\documentclass[twocolumn, amssymb,nobibnotes,aps,prb,nopacs]{revtex4}

\usepackage{amsmath}
\usepackage{graphicx}
\usepackage{subfig}
\newcommand \dd[1]  { \,\textrm d{#1} }   
\usepackage[justification=RaggedRight,font=footnotesize]{caption}

\usepackage{multirow}

\begin{document}

\title{Study of trapping effect on ion-acoustic solitary waves  \\ based on a fully kinetic simulation approach}

\author{S. M. Hosseini Jenab\footnote{Email: Mehdi.Jenab@nwu.ac.za}}
\author{F. Spanier \footnote{Email: Felix@fspanier.de}}
\affiliation{Centre for Space Research, North-West University, Potchefstroom Campus, Private Bag X6001, Potchefstroom 2520, South Africa}

\date{\today}

\begin{abstract}
A fully kinetic simulation approach, treating each plasma component based on the Vlasov equation,
is adopted to study the disintegration of an initial density perturbation (IDP)
into a number of ion-acoustic solitary waves (IASWs) in the presence of the trapping effect of electrons. 
The nonlinear fluid theory developed by Schamel \cite{schamel_1,schamel_2,schamel_3,schamel_4,schamel_5} has identified 
three separate regimes of ion-acoustic solitary waves based on the trapping parameter.
Here, the disintegration process and 
the resulting self-consistent IASWs are studied in
a wide range of trapping parameters covering all the three regimes continuously.
The dependency of features such as the time of disintegration,  the number,  speed and size of IASWs
on the trapping parameter are focused upon.
It is shown that an increase in this parameter slows down 
the propagation of IASWs while decreases their sizes in the phase space.
These features of IASWs tend to saturate for large value of trapping parameters. 
The disintegration time shows a more complicated behavior than what was predicted by the theoretical approach.
Also for the case of trapping parameters bigger than one, 
propagation of IASWs is observed in contrast with the theoretical predictions.
The kinetic simulation results unveil a smooth and well-defined dependency of solitary waves' features on the trapping parameter,
showing the possibility of bridging all the three regimes.
Finally, it is shown that for $\beta$ around zero,
the electron phase space structure of 
the accompanying vortex stays symmetric.
The effect of the electron-to-ion temperature ratio  
on the disintegration and the propagation of IASWs are considered 
as a benchmarking test of the simulation code (in the nonlinear regime).
\end{abstract}
\maketitle

\section{Introduction}
A solitary wave is a localized nonlinear wave which propagates unaltered 
due to an exact balance between the widening tendency of dispersive effects 
and the steepening inclination of nonlinear effects.
Two characteristics are associated with these waves,
 a) propagation without any changes in their features such as velocity and shape (height and width),
 b)stability against mutual collisions.
The first condition describes a \textit{solitary wave}.
A localized wave satisfying both conditions is called \textit{soliton},
the suffix ``on'' is used to indicate the particle property \cite{Wadati2001841}. 
Mathematically, these attributes arise from the fact that solitons are among the exact solutions of
the so called \textit{integrable nonlinear partial differential equations}.
These equations support infinite number of conservation laws, hence called \textit{integrable}.
Qualitatively speaking, the infinite number of conservation laws guarantees that the solutions of these equations 
should remain the same in time and during mutual collisions 
by imposing restrict conditions on their existence \cite{Debnath20071003}.

Different solitary waves have been predicted in plasmas,
especially in multi-species plasmas which are able to support complicated shapes
\cite{Verheest2013,bharuthram1992large,verheest1988ion,sultana2015electron,verheest2016modified},
and numerous observations and experiments have approved some of such nonlinear structures \cite{nakamura1985experiments,ludwig1984observation,
lonngren1998ion,ikezi1970formation,cooney1991experiments}. 
One of the best examples of the existence of solitary waves in plasma includes electrostatic solitary waves (ESWs) in the broadband electrostatic noise (BEN) 
observed by different satellites(e.g., Polar\cite{franz1998polar}, GEOTAIL \cite{matsumoto1994electrostatic,kojima1997geotail}, 
FAST \cite{catte1151998fast}, and Cluster \cite{pickett2003solitary,hobara2008cluster,pickett2004solitary,kakad2016slow})
in various regions of the Earth's magnetosphere.

Ion-acoustic solitary waves (IASWs), the main solitary waves in plasma physics, 
exist in the simplest form of multi-species plasmas;
i.e. hot electrons and cold ions. 
IASWs are the first solitary waves to be discovered in plasma physics.
By the seminal work of Washimi and Taniuti \cite{Washimi1966996}, the governing nonlinear fluid equations of this plasma
are reduced to KdV equation employing reductive perturbation technique. 
This model assumes isothermal and adiabatic electrons and solves fluid dynamics equations for ions coupled with Poisson's equation.
A critical Mach number (speed of soliton) has been established by this model below which IASWs can't exist\cite{baluku2010new}.
However, this model ignores the trapping mechanism and its effect on the speed and shape of IASWs. 

Trapped particles involve particles resonating with the potential well of a solitary wave and accompanying it in 
its propagation. They appear as a hole or hump in the phase space.
Schamel \cite{schamel_3} has included the effect of trapped electrons,
and showed that it introduces its own nonlinearity term in KdV equation (hence modified KdV).
This nonlinearity depends on the number of trapped electrons 
(parametrized by $\beta$ called trapping parameter).
By comparing this trapping nonlinearity with the usual nonlinearity,
three different IASWs are possible,
including the normal solitary waves; i.e. KdV solutions.

The question of physical importance about the temporal evolution of solitary waves includes the so called
\textit{disintegration/breaking}.
Mathematically speaking, N-soliton solutions 
have been achieved for KdV and KdV-type equations through multiple approaches such as inverse scattering transform (IST) \cite{Gardner19671095},
Hirota's method\cite{Hirota1971,Hirota1972} and exp-function method \cite{lee2011exact,wu2013new,taha2013new}. 
Hence it can generally be assumed that any nonlinear structure will break into solitary waves given enough time.
Zabusky \cite{Zabusky1965}, for the first time, has shown  
the disintegration of a nonlinear structure into a few solitons through simulation for collisionless plasmas.

Two different simulation approaches can be employed to address the problem of disintegration. 
Self-consistent solitary waves produced by this process can also be used to indirectly test 
the validity of the nonlinear fluid theory (Sagdeev's solutions). 
Each of these simulation methods includes 
following a dynamical equation coupled with Poisson's equation (electrostatic limit).
The first type of simulation methods focuses on the fluid-type quantity (i.e. density).
They comprise two major methods:
a) KdV method which is based on KdV or KdV-type equations, 
b)fluid methods which use the multi-fluid equations of continuity, momentum, and energy of each species.
This type of simulation approaches is unable to fully incorporate the trapping effect,
since much of the details of such process happen in velocity direction in phase space. 
Even by starting from a distribution function with a hole
-hence with an exact initially perturbed density (Schamel's method \cite{schamel_4})-
the temporal evolution of such a hole is ignored during simulation in phase space. 
Kakad \textit{et al.} \cite{Kakad2013} studied the disintegration progress (chain formation)  
mostly focusing on ESWs in BEN 
based on a multi-species fluid model \cite{lakhina2009mechanism}.
These simulations revealed that a stationary IDP would break into 
two oppositely propagating identical ion-acoustic 
solitons/soliton and (Langmuir and ion-acoustic) wavepackets for the case of small/large IDP.

The second simulation approach refers to the so called the \textit{kinetic simulation} 
which contains two methods: PIC and Vlasov.
Particle in cell (PIC) employs the concept of super-particles and follows their dynamics.
There have been a few notable PIC simulation attempts to study IASWs and the disintegration process.
These simulations have tried to address the question about the validity of the fluid theory assumptions. 
However, since the PIC method doesn't deal with the concept of distribution function directly at each time step,
its dynamic is often ignored in PIC reports.
Furthermore, details of distribution function evolution can not be traced, due to the inherent noise in PIC, 
especially the hole/hump accompanying solitary waves.
Sharma \textit{et al.} \cite{Sharma2015} have carried out PIC simulations to study the large amplitude IASWs.
Relative agreement was reported between simulation results and the solutions obtained via Sagdeev's method.
The existence of an upper limit for the amplitude of IASWs has been reported using a PIC code \cite{Qi20153815}. 
However, they all have carried out their studies ignoring the trapping process and treated electrons 
as isothermal background (Boltzmann's distribution is used for electrons). 
Therefore, the effect of trapping and the deviation it causes (on evolution and features of the nonlinear solutions) 
is missing in these simulations.
Kakad \textit{et al.} \cite{Kakad20145589} reported that PIC and fluid simulation are in close agreement for small amplitude IDPs.
Discrepancies for large amplitude IDPs is shown to exist between PIC and fluid simulations results 
when electron dynamics (the trapping effect of electrons) is included.

In this study, we employed a fully kinetic (Vlasov) approach 
\cite{jenab2014vlasov,jenab2014multicomponent,jenab2011preventing}
to study the disintegration of IDPs into a number of IASWs for the first time.
All the plasma components, namely electrons and ions, are treated based on the Vlasov equations in this approach.
In order to insert an IDP into simulation, we have utilized Schamel distribution function (Eq. \ref{Schamel_Dif}). 
The dynamics of disintegration can be followed in the phase space and therefore the restrictions in fluid or KdV methods are removed completely. 
Due to the low noise, the nonlinear structures accompanying IASWs in the phase space are rigorously traced and reported here 
(which was missing in the PIC simulations).
This model enables us to provide a comprehensive view on the dynamics in order to address the questions 
about the validity of nonlinear fluid models \cite{schamel_3,schamel_4,schamel_5}.

\section{Basic Equations and Numerical Scheme} \label{B_equations}
A brief review of basic assumptions and approximations leading to the three different regimes will be presented here
(for details see \cite{schamel_1,schamel_2,schamel_3,schamel_4,schamel_5}).
It has been shown that the following distribution function 
(called Schamel distribution function) satisfies the continuity and positiveness conditions
while producing a hole in its phase space \cite{schamel_1,schamel_2}:
\begingroup\makeatletter\def\f@size{8.3}\check@mathfonts
\def\maketag@@@#1{\hbox{\m@th\large\normalfont#1}}%
\begin{equation*}
f_{s}(v) =  
  \left\{\begin{array}{lr}
     A \ exp \Big[- \big(\sqrt{\frac{\xi_s}{2}} v_0 + \sqrt{E(v)} \big)^2 \Big]   &\textrm{if}
      \left\{\begin{array}{lr}
      v<v_0 - \sqrt{\frac{2E_{\phi}}{m_s}}\\
      v>v_0+\sqrt{\frac{2E_{\phi}}{m_s}} 
      \end{array}\right. \\
     A \ exp \Big[- \big(\frac{\xi_s}{2} v_0^2 + \beta_s E(v) \big) \Big] &\textrm{if}  
     \left\{\begin{array}{lr}
      v>v_0-\sqrt{\frac{2E_{\phi}}{m_s}} \\
      v<v_0 + \sqrt{\frac{2E_{\phi}}{m_s}} 
      \end{array}\right.
\end{array}\right.
\label{Schamel_Dif}
\end{equation*}\endgroup
in which $A = \sqrt{ \frac{\xi_s}{2 \pi}} n_{0s}$,
and $\xi_s = \frac{m_s}{T_s}$ are amplitude and the normalization factor respectively.
$E(v) = \frac{\xi_s}{2}(v-v_0)^2 + \phi\frac{1}{T_s q_s}$ represents the (normalized) energy of particles.
$v_0$ stands for the velocity of the solitary wave.
In the set of the simulations presented here,
this distribution function has been used to introduce a stationary IDP ($v_0 =0$) at $x_0$:
\begin{equation}
 \phi = \psi \ exp (\frac{x-x_0}{\Delta})^2.
\end{equation}

To simplify the equations, all variables have been rescaled into dimensionless forms.
Space and time are normalized by $\lambda_{Di}$ and $\omega^{-1}_{pi}$ respectively,
where $\omega_{pi}  = \sqrt{n_{i0} e^2 /(m_i \epsilon_0) }$ denotes the ion plasma frequency
and $\lambda_{Di} = \sqrt{\epsilon_0 K_B T_i /(n_{i0} e^2) }$ is the characteristic
ion Debye length.
The velocity variable $v$ has been scaled by
the ion thermal speed $v_{th_i} = \sqrt{K_B T_i/m_i}$,
while the electric
field and the electric potential have been reduced by $K_B T_i/(e \lambda_{Di})$
and $K_B T_i/e$ respectively (here, $K_B$ is Boltzmann's constant).
The densities of the two species are normalized by $n_{i0}$, while energy is scaled by $K_B T_i$.
Note that by this normalization, ion sound velocity and electron plasma frequency are 
$v_C = 8.06$ and $\omega_{pe} = 10$ respectively.

\begin{figure}[htp]
  \subfloat{\includegraphics[width=0.35\textwidth]{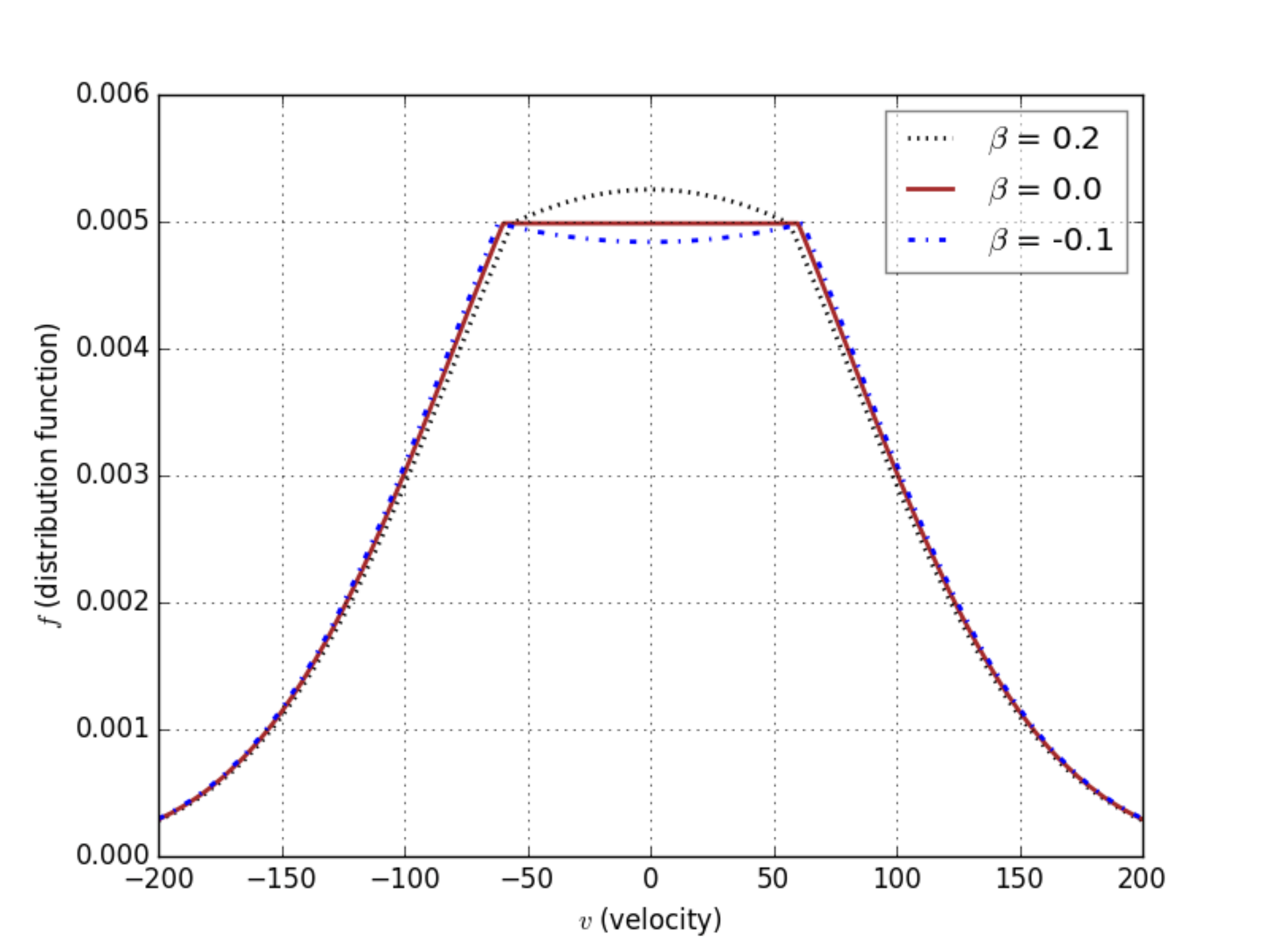}}
  \caption{Trapped electrons distribution function around $v_0=0$ 
  appears as a hole ($\beta<0$), a plateau ($\beta=0$) and a hump($\beta>0$).}
  \label{DF_Beta}
\end{figure}
$\beta$ is the so called \textit{trapping parameters} which describes the distribution function around $v_0$. 
Based on $\beta$, Fig.\ref{DF_Beta} shows that the distribution function of trapped particles 
can take three different types of shapes, 
namely \textit{hole} ($\beta<0$), \textit{plateau} ($\beta = 0$) and \textit{hump} ($\beta>0$).
Considering the temperature of trapped particles $\beta = \frac{T_{f}}{T_{t}}$  
can be defined as the temperature ratio of free ($T_f$) to trapped ($T_t$) particles.
As $T_t \longrightarrow \pm \infty$, the distribution function's shape of the trapped particles widens 
and goes to a plateau ($\beta = 0$) in the trapping region. 
A hole in a distribution function,
which looks like a dip in velocity direction,
is associated to a negative temperature.
Note that although a distribution function can not take the shape of a dip 
in the velocity direction (violation of the positiveness of distribution function),
a part of distribution function is allowed to form a dip as far as it stays above zero.

By integrating the distribution function on the velocity direction,
one can obtain the densities of (two) plasma constituents.
Furthermore by applying 
the \textit{necessary conditions} such as current condition and conditions on classical potential,
\textit{nonlinear dispersion relation}(NDR) can be achieved. 
NDR contains the necessary information to build solitary solutions.
Moreover, Washimi and Taniuti \cite{Washimi1966996} approach -reductive perturbation technique-
can be extended to involve the trapping effect \cite{schamel_3,schamel_4}.
The resulting dynamical equation takes the form:
\begin{equation}
  \frac{\partial \phi}{\partial t} +
  \big(1+\Omega(\phi) \big) \frac{\partial \phi}{\partial x} +
  \frac{1}{2} \frac{\partial^3 \phi}{\partial x^3 }  = 0,
  \label{Equ_genKdV}
\end{equation}
in which $\Omega(\phi)$
-the nonlinearity coefficient-
accepts three different forms.
Each form is allocated to a specific range of $\beta$ with its own ion-acoustic solitary solutions.
Firstly,
for $|b| \ll O(\sqrt{\psi})$
(in which $b = \frac{1-\beta}{\sqrt{\pi}}$)
the nonlinearity coefficient is in the form of $\Omega = \phi$.
In this regime,
the trapping nonlinearity is negligible,
and Eq.\ref{Equ_genKdV} takes the form of the well-known KdV equation.
Secondly,
for $|b| \approx O(\sqrt{\psi})$,
Eq.\ref{Equ_genKdV} is called Schamel-KdV equation with 
$\Omega = b \sqrt{\phi} + \phi$.
Here,
the strength of the trapping nonlinearity 
is in the same order as the usual nonlinearity.
Finally,
when the trapping nonlinearity is dominant ($|b| \gg O(\sqrt{\psi})$),
the usual nonlinearity can be ignored.
Therefore the nonlinearity coefficient is $\Omega = b \sqrt{\phi} $ -Schamel regime.
However,
one should note that 
the method and the results mentioned above
are restricted with the two following conditions:
1) the small amplitude approximation and 
2) electrons acting as quasi-stationary.

The simulation method,
employed here,
has been developed by the authors based on the method called \textit{Vlasov-Hybrid Simulation}(VHS),
which was initially proposed by Nunn \cite{nunn1993novel} 
(for details see \cite{jenab2014vlasov,jenab2014multicomponent,jenab2011preventing}).
It follows the trajectories of the 
so called \textit{phase points} \cite{kazeminezhad2003vlasov} in the phase space,
depending on Liouville's theorem as the theoretical framework.
Preserving entropy $(\int f  \ln f  \dd v \dd x)$, entropy-type quantities $(\int f^2  \dd v \dd x)$,
and energy stands as one of the major advantage of the method.
In simulations presented in this paper,
each plasma species (i.e. electrons and ions)
is described by the (scaled) Vlasov equation:
\begin{multline}
\frac{\partial f_s(x,v,t)}{\partial t} 
+ v \frac{\partial f_s(x,v,t)}{\partial x} 
\\ +  \frac{q_s}{m_s} E(x,t) \frac{\partial f_s(x,v,t)}{\partial v} 
= 0, \ \ \  s = i,e
\label{Vlasov}
\end{multline}
where $s = i,e$ represents the corresponding plasma species.
The variable $v$ denotes velocity in (1D) phase space.
$q_s$ and $m_s$ are normalized by $e$ and $m_i$ respectively.
Densities of the plasma components are given upon integration as:
\begin{equation}
n_s(x,t) = n_{0s}\int f_s(x,v,t) dv
\label{density}
\end{equation}
which are coupled with Poisson's equation:
\begin{equation}
\frac{\partial^2 \phi(x,t)}{\partial x^2}  = n_e(x,t) - n_i(x,t)
\label{Poisson}
\end{equation}
The equilibrium values $n_{s0}$ are assumed to satisfy 
the quasi-neutrality condition ($n_{e0} = n_{i0} $) at the initial step of simulations.

Our numerical procedure is as follow.
At each time step,
the distribution functions are calculated from the Vlasov equations (Eq. \ref{Vlasov}).
Then, the number density of each plasma species
is obtained by integration (Eq. \ref{density}).
By substituting the corresponding densities into Poisson's equation (Eq. \ref{Poisson}), 
the electric field is obtained. 
The electric field is
then put into the Vlasov equations.
This cycle is iterated, and the results
are retained at each step.
Energy and entropy preservation is meticulously observed during all the simulations to stay below one percent deviation.

The constant parameters which remain fixed through all
of our simulations are: 
$\frac{m_i}{m_e} = 100$, time step $d\tau = 0.01$, $L = 4096$, 
where L is the length of the simulation box.
$\psi = 0.2$ and  $\Delta = 500$ are the amplitude and width of the stationary IDP respectively.
These values introduce a large IDP into the simulation domain which creates at least two IASWs which we need for the analysis
$\theta = \frac{T_e}{T_i} = 1,5$ are chosen for Sec. \ref{theta}. 
The values of $\beta$ were modified between
successive simulations in range of $-0.5\leq \beta \leq 10$, while $\theta = 64$ for Sec \ref{beta}.
We have considered a two-dimensional phase space with one spatial and one velocity axis.
The phase space grid $(N_x, N_v)$ size is $(4096, 4000)$.
The periodic boundary condition is employed on x-direction.

\section{Results and Discussion} \label{Results}
The dynamical progression following the initial step of a stationary IDP can be divided into two chronological steps. 
Early in the temporal evolution,
the stationary IDP will break into two oppositely 
propagating IDPs due to the symmetry of 
the distribution function in the velocity direction.
\begin{figure}[htp]
  \subfloat[electron distribution function]{\includegraphics[width=0.25\textwidth]{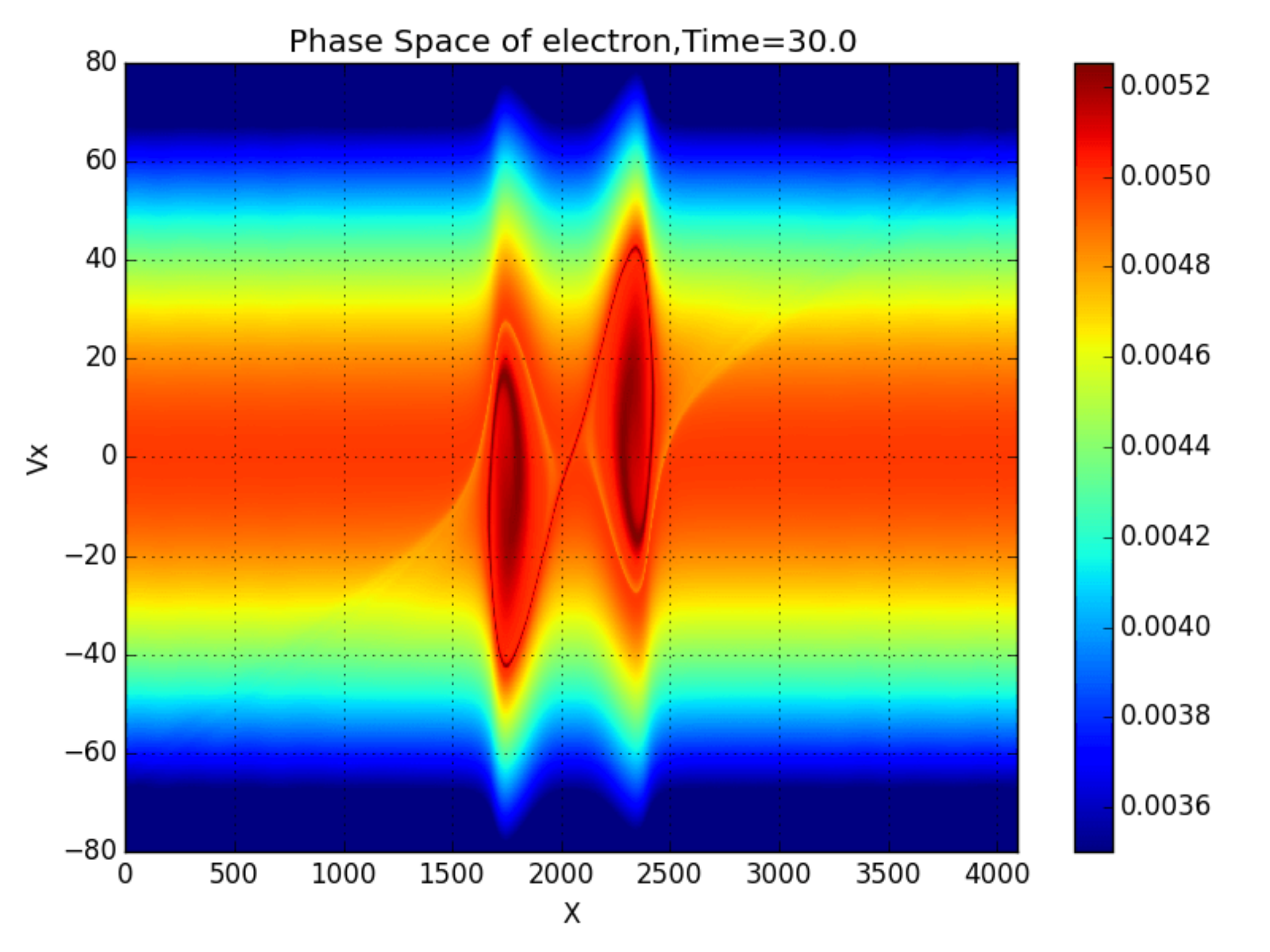}\label{Breaking_IDP_1}}
  \subfloat[number density]{\includegraphics[width=0.25\textwidth]{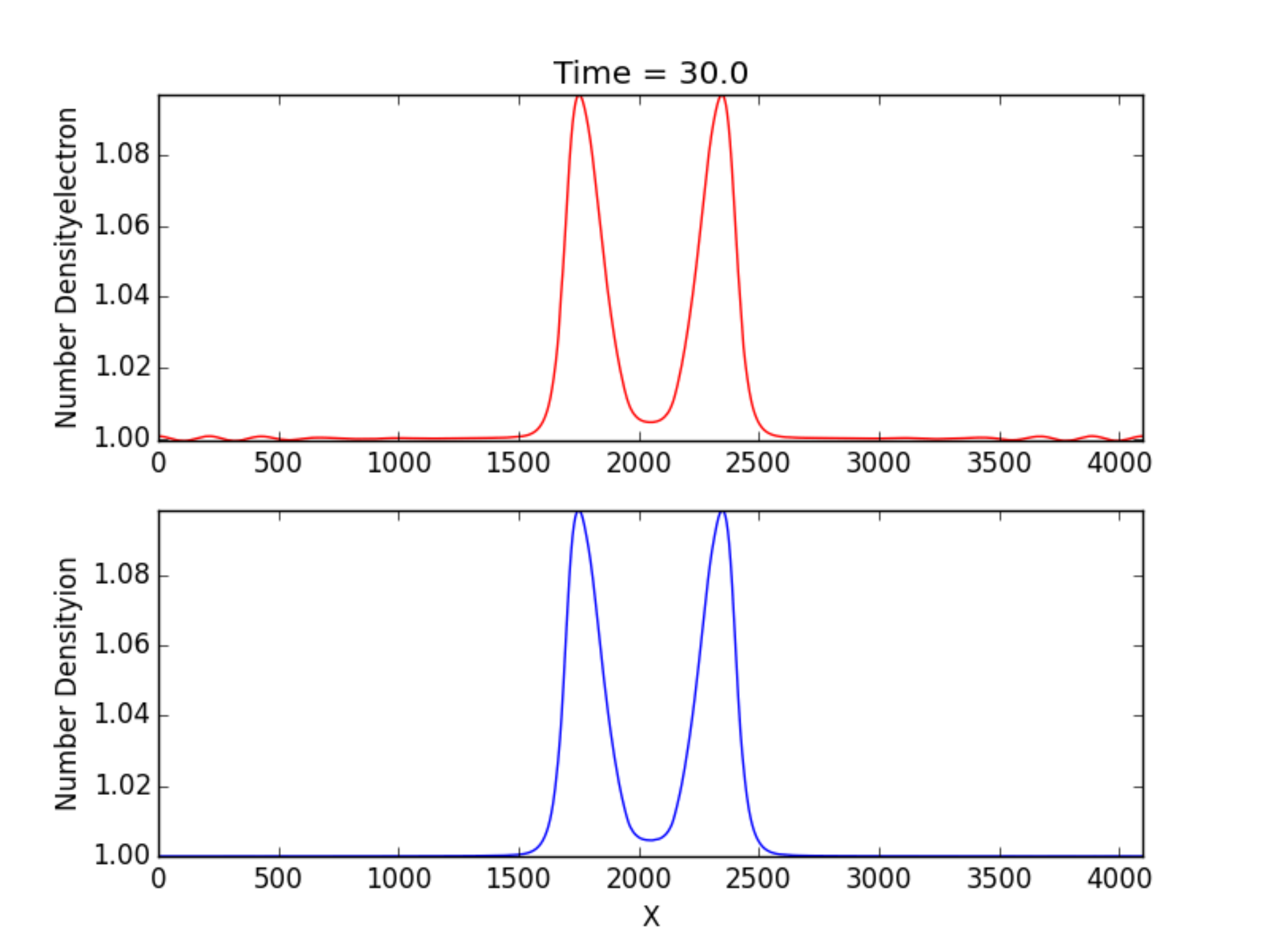}\label{Breaking_IDP_2}} \\
  \subfloat[electron distribution function]{ \includegraphics[width=0.25\textwidth]{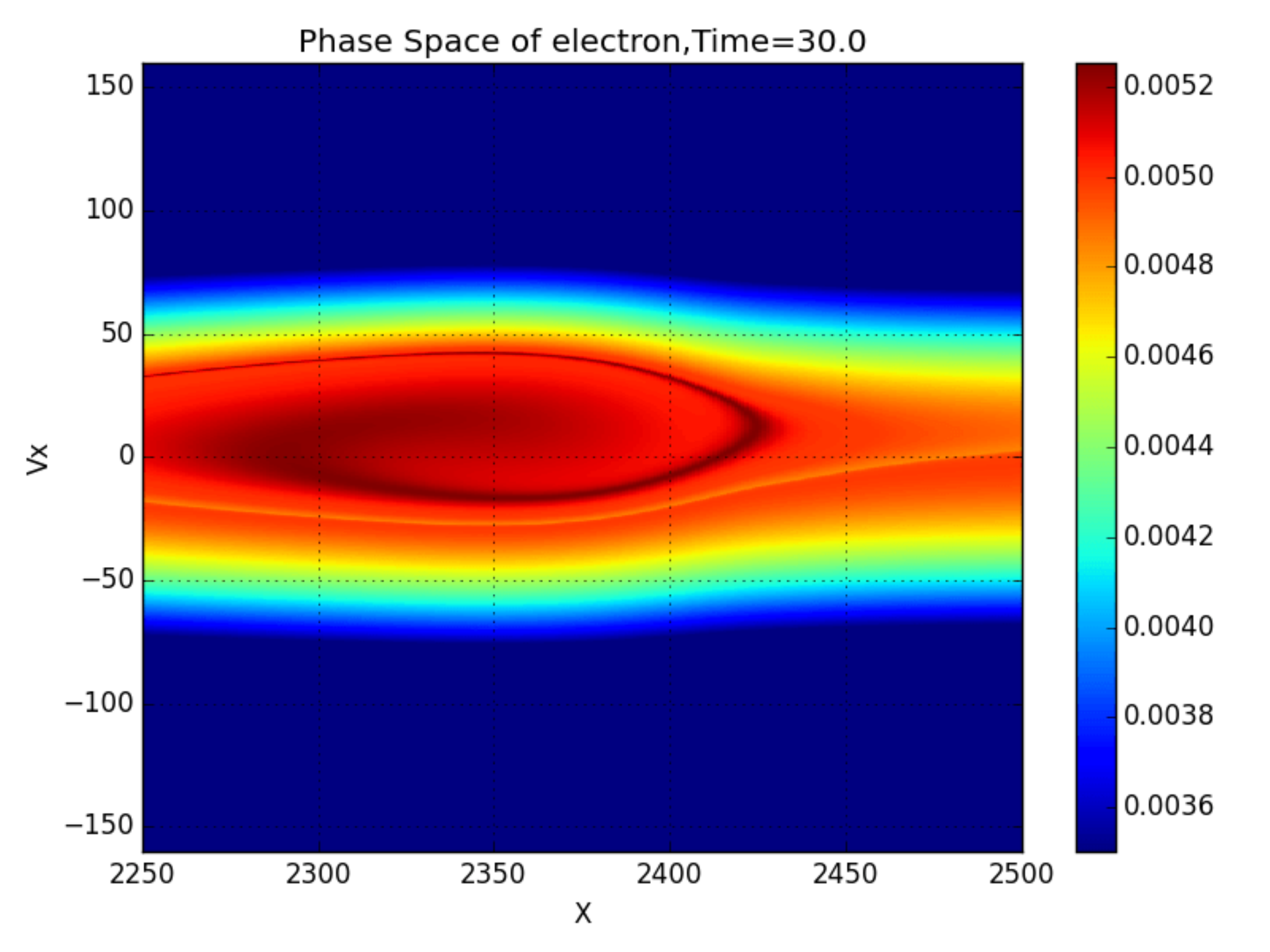}\label{Breaking_IDP_3}}
  \subfloat[ion distribution function]{ \includegraphics[width=0.25\textwidth]{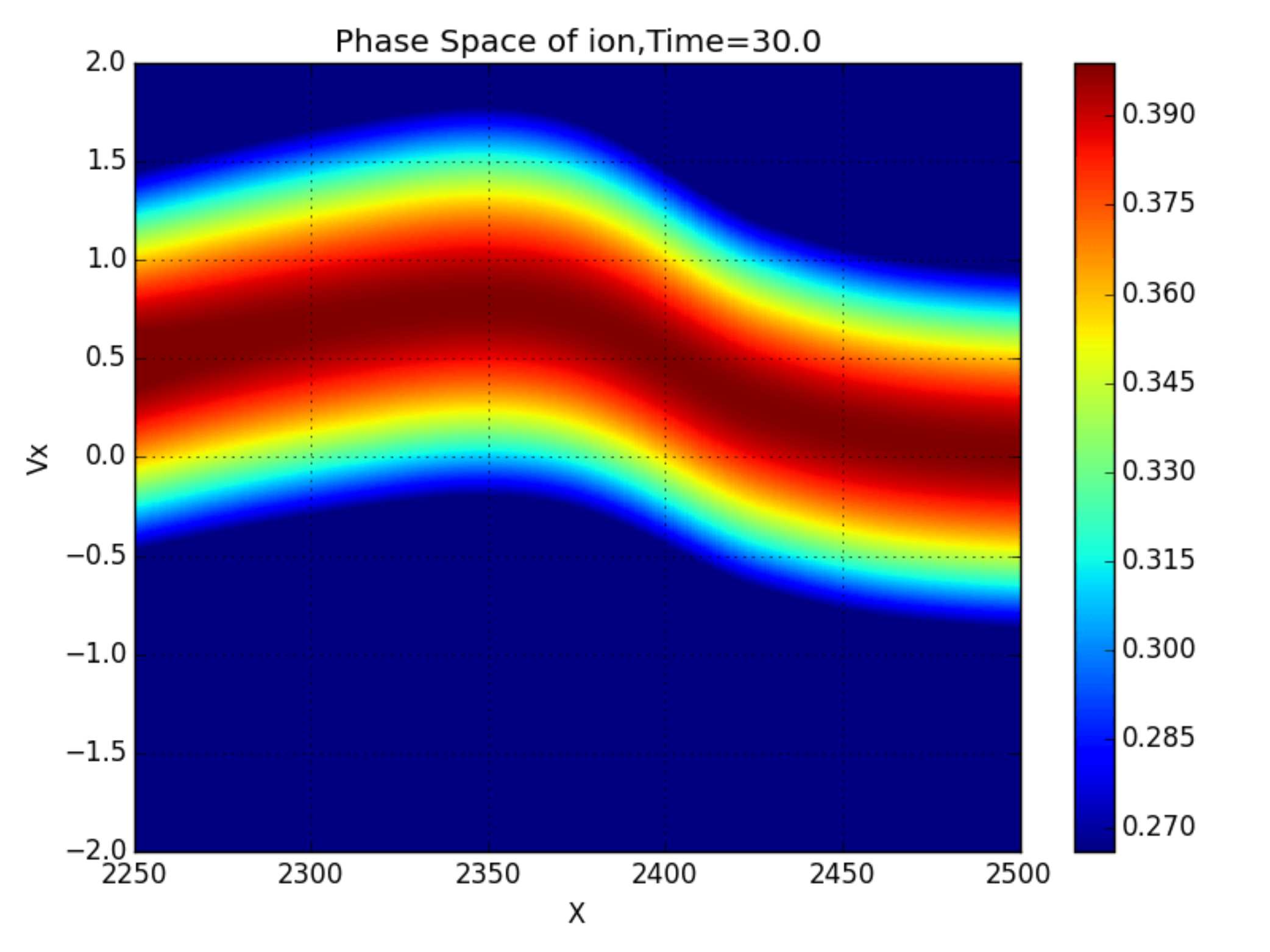}\label{Breaking_IDP_4}} 
  \caption{ The initial break-up of a stationary IDP into two moving IDPs is presented
  for the case 4 of table \ref{table} with $\beta =0.2$ and $\theta = 64$ at time $\tau = 30$.
  The electron distribution function (\ref{Breaking_IDP_1}), 
  the number densities of ions and electrons(\ref{Breaking_IDP_2}) are shown.
  Steepening on the propagation side because of the nonlinearity can be observed.
  The vortex-shape hump in the electron distribution function (\ref{Breaking_IDP_3})
  suggests the existence of the trapping effect,
  while ions don't show any trapping (\ref{Breaking_IDP_4}).}
  \label{Breaking_IDP}
\end{figure}
This breaking happens (in all the cases presented here) much faster 
than the disintegration (of moving IDPs into IASWs) depending on the $\theta$.
For $\theta = 64$, it starts immediately after the initial step $(\tau<5)$ 
(compared to the long time simulation $\tau = 200$). 
The velocities of the moving IDPs are self-consistent
and depend on the trapping parameter $\beta$.
Fig. \ref{Breaking_IDP} shows a hump representing the trapped electrons for each of the moving IDPs.
However, there is no hole/hump around the velocity of the moving IDPs in the ion distribution function, suggesting that ion trapping doesn't exist. 
Therefore, the set of simulations presented here shows the disintegration in the presence of electron trapping without ion trapping.

\begin{figure}[htp]
  \subfloat[Langmuir wavepacket]{\includegraphics[width=0.25\textwidth]{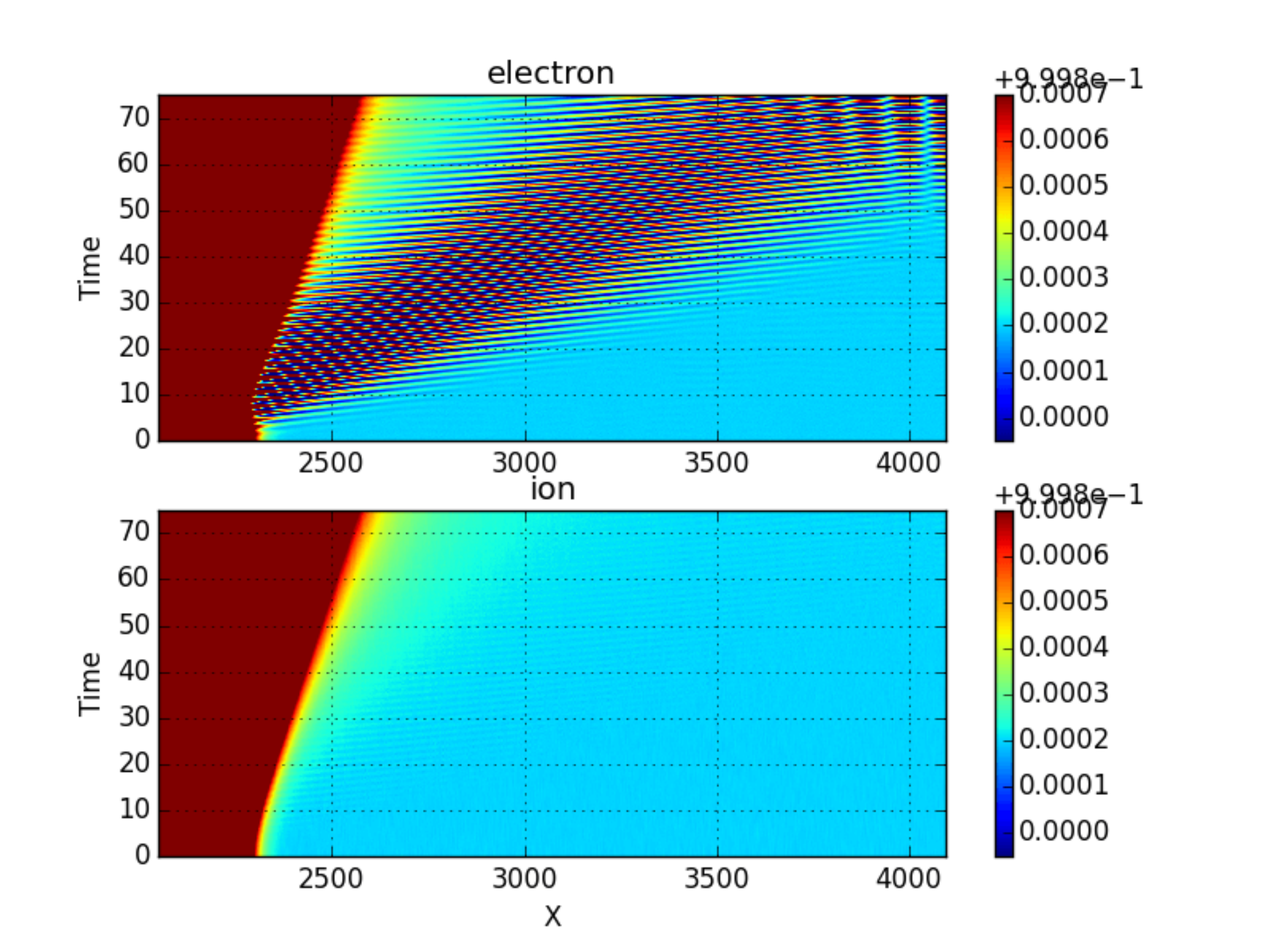}\label{WavePacket_Den_1}}
  \subfloat[ion-acoustic wavepacket]{\includegraphics[width=0.25\textwidth]{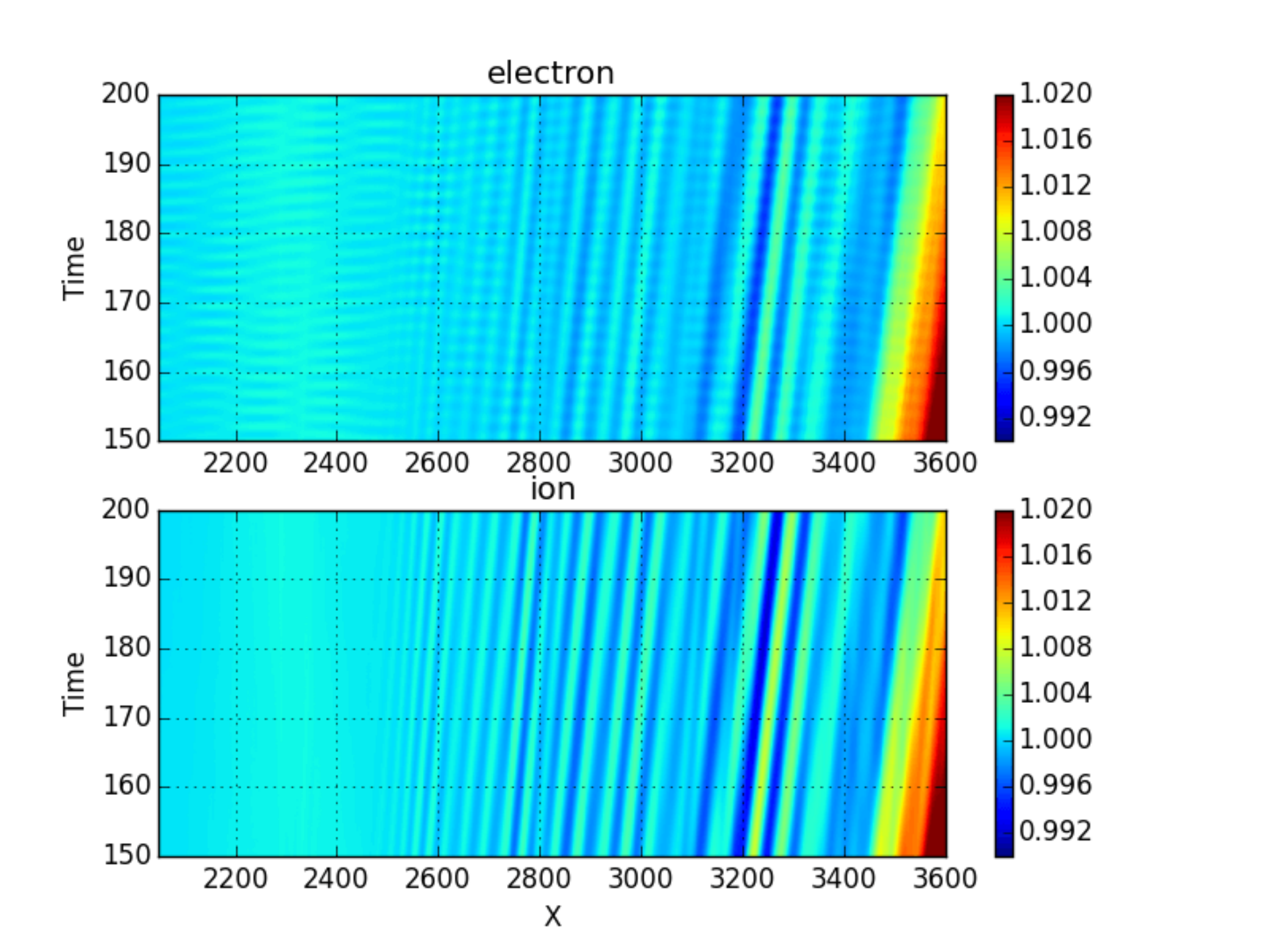}\label{WavePacket_Den_2}}
  \caption{ Propagation of wavepackets for the case 4 of table \ref{table} is shown with $\beta =0.2$ and $\theta = 64$.
  The Langmuir wavepacket propagates faster than IASWs and hence stays ahead of them (\ref{WavePacket_Den_1}). 
  The ion-acoustic wavepacket propagates slower than IASWs, and therefore can be witnessed behind them (\ref{WavePacket_Den_2}). }
  \label{WavePacket_Den}
\end{figure}

At the later stage,
each of the moving IDPs starts steepening on the propagation side
due to the nonlinearity effects. 
Next,
they disintegrate into a number of IASWs 
and two wavepackets (including Langmuir waves and ion-acoustic waves).
Solitary waves will be aligned based on their amplitude,
since the higher the amplitude, the faster the propagation speed.
Splitting of the solitary waves will happen sequentially.
initially, the first solitary wave will emerge from the moving IDP.
The remaining part of IDP might break into more solitary waves
in the same manner that the first solitary wave emerges. 
Otherwise, this part will turn into the second solitary wave.
The Langmuir wavepacket propagates much faster than IASWs,
and appears ahead of them.
On the other hand, ion-acoustic wavepacket can be observed behind IASWs since they are slower \cite{schamel_4,Kakad2013}.
Fig. \ref{WavePacket_Den} shows the propagation of both wavepackets.
Widening of the Langmuir wavepacket can be observed due to the dispersive effect.
These wavepackets have also been witnessed in fluid simulations \cite{Kakad2013} and PIC simulations \cite{Kakad20145589}.

\subsection{Effect of Ion-to-electron Temperature Ratio ($\theta$) } \label{theta}
Theoretically,
it is shown that for $\theta = \frac{T_e}{T_i}<3.5$,
no IASWs can propagate\cite{schamel_3}.
Firstly,
we consider the case $\theta =1 $, 
and the kinetic simulation approach shows that 
the moving IDPs can't disintegrate into IASW.
Instead they widen and weaken down to the noise level.  
\begin{figure}[htp]
   \centering
  \subfloat[number density for $T_e = T_i$]{\includegraphics[width=0.25\textwidth]{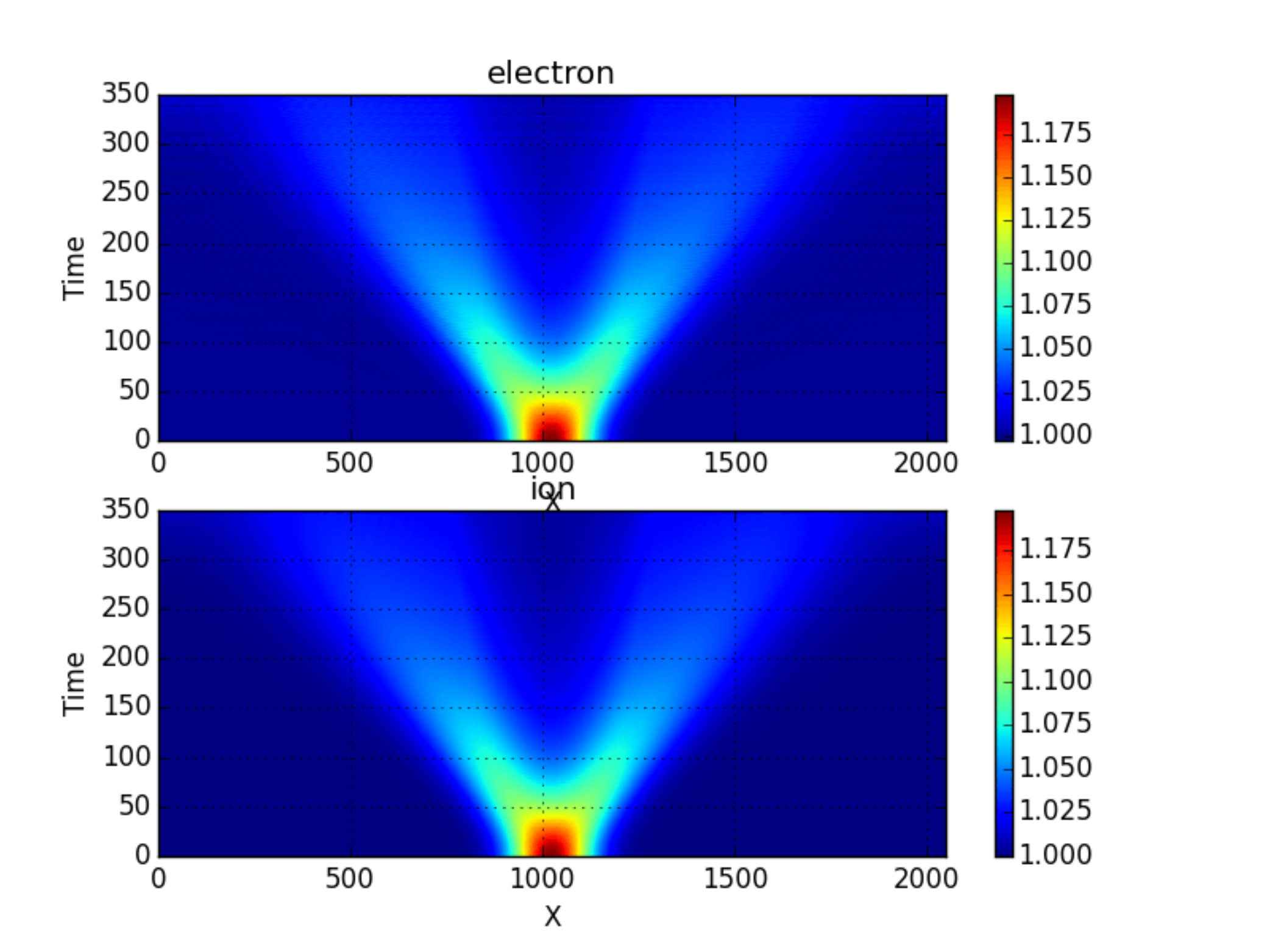}\label{T_eff_Den_1}}
  \subfloat[number density for $T_e = 5 T_i$]{\includegraphics[width=0.25\textwidth]{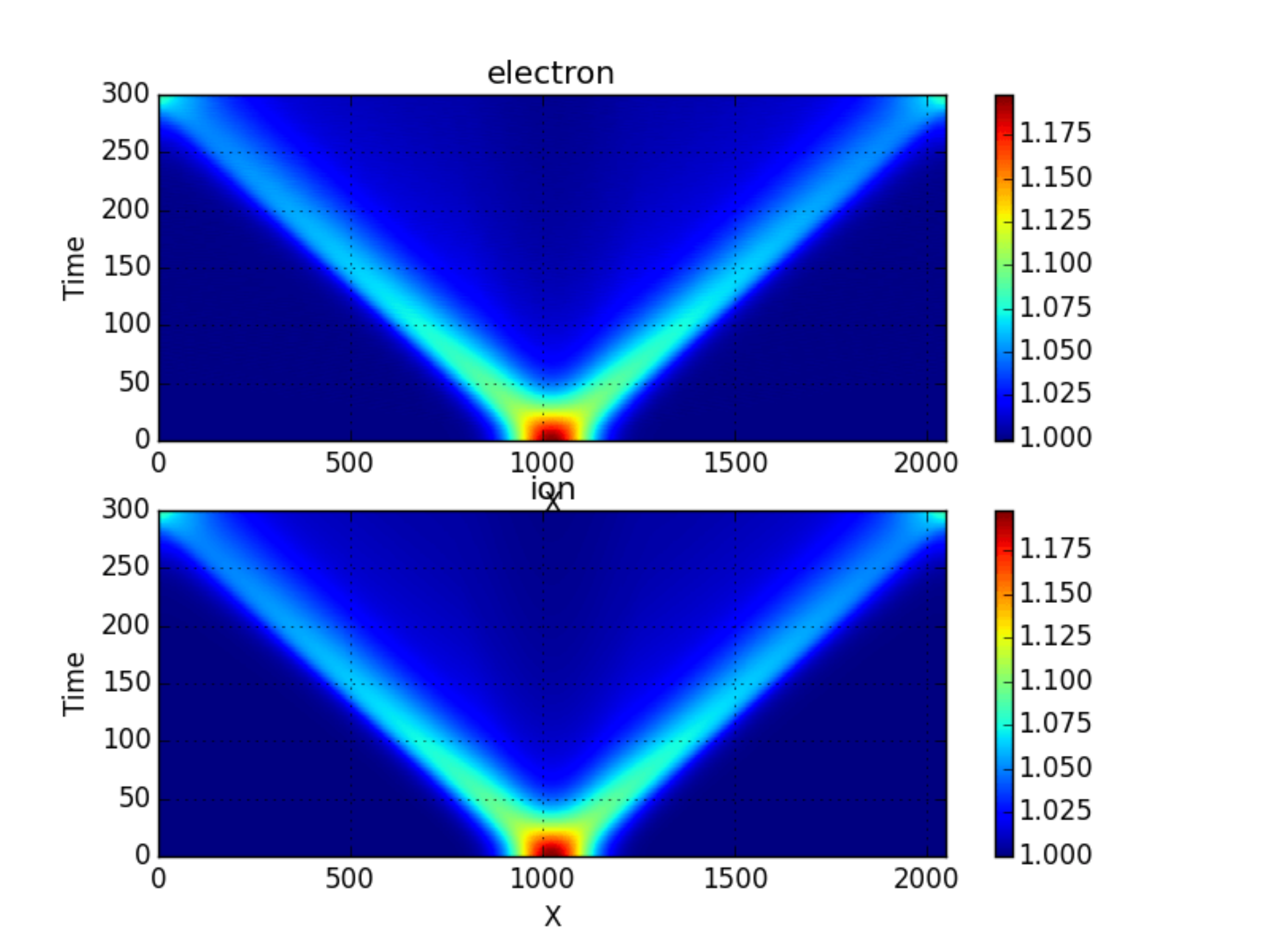}\label{T_eff_Den_2}}
  \caption{The effect of electron-to-ion temperature ratio on the propagation of solitary waves
  is presented for two different temperature ratios.
  In case of $T_e = T_i$, ion-acoustic solitary waves can't propagate and hence the two moving IDPs can't disintegrate into ion-acoustic solitary waves.
  Instead these IDPs widens and disappears (\ref{T_eff_Den_1}).
  In case of $T_e = 5T_i$ the IDPs disintegrate into a solitary wave and Langmuir/IA wavepackets. 
  The propagation of two oppositely moving solitary waves with velocity $\pm 2.75$ can be observed (\ref{T_eff_Den_2}). }
  \label{T_eff_Den}
\end{figure}
Fig. \ref{T_eff_Den} presents the results of two cases $\theta = 1 $ and $\theta = 5 $.
The widening of moving IDPs can be observed in case of $\theta = 1 $,
while the propagation of ion-acoustic solitary waves exists in case of  $\theta = 5 $.

\begin{figure}[htp]
  \centering
  \subfloat[FFT for $T_e =  T_i$]{\includegraphics[width=0.25\textwidth]{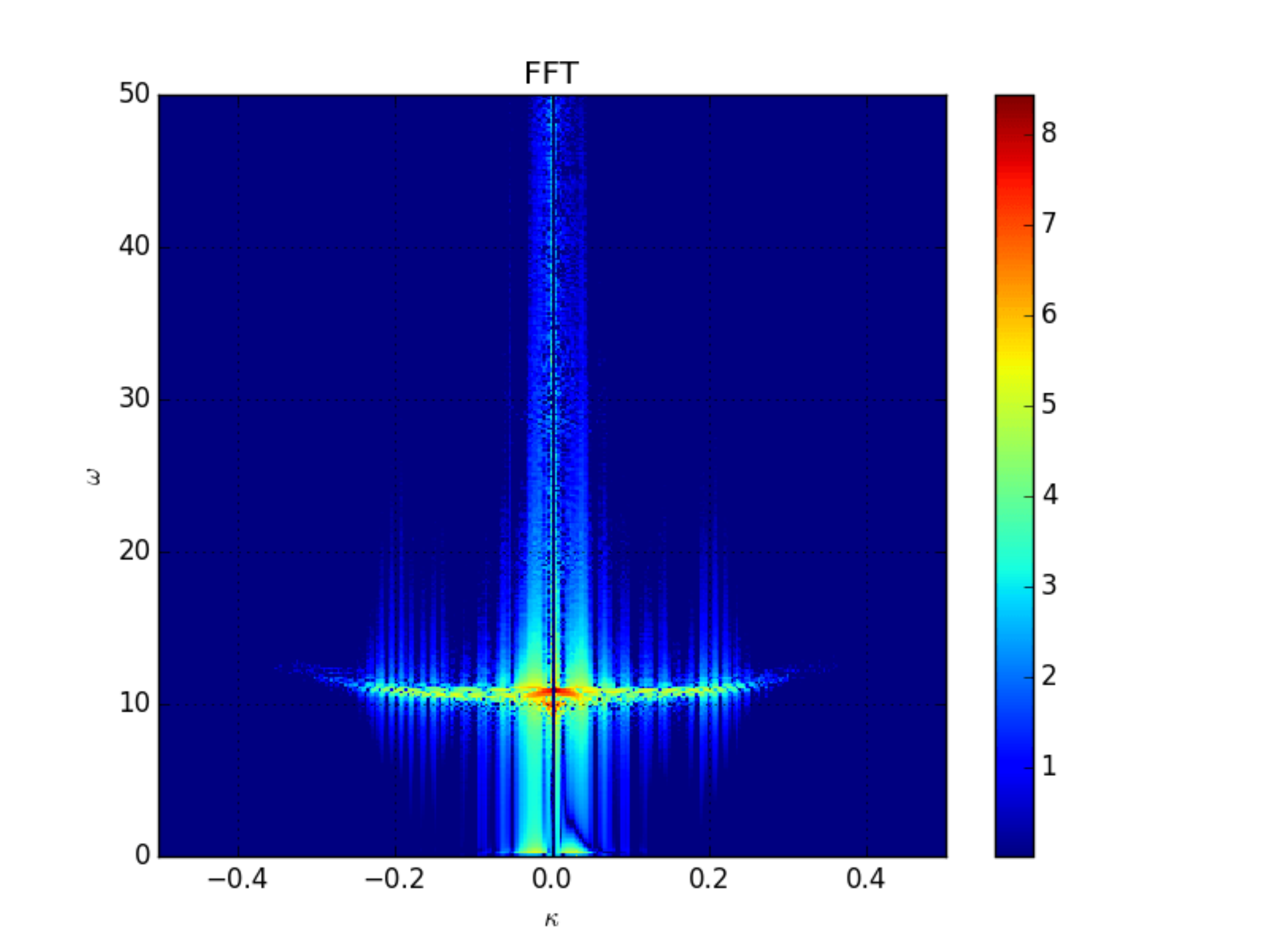}
				  \includegraphics[width=0.25\textwidth]{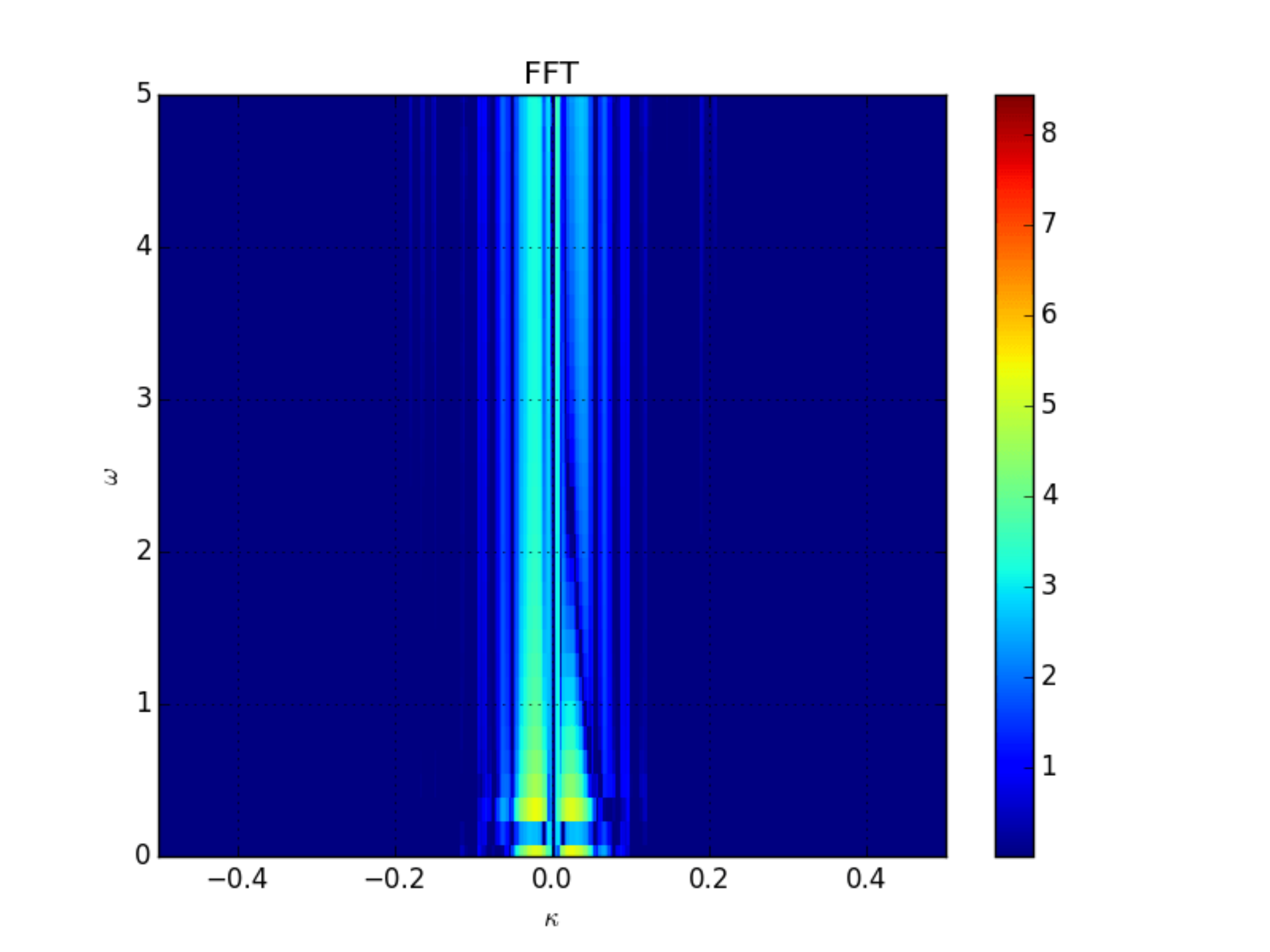}
  \label{T_FFT_1}} \\
  \subfloat[FFT for $T_e = 5 T_i$]{\includegraphics[width=0.25\textwidth]{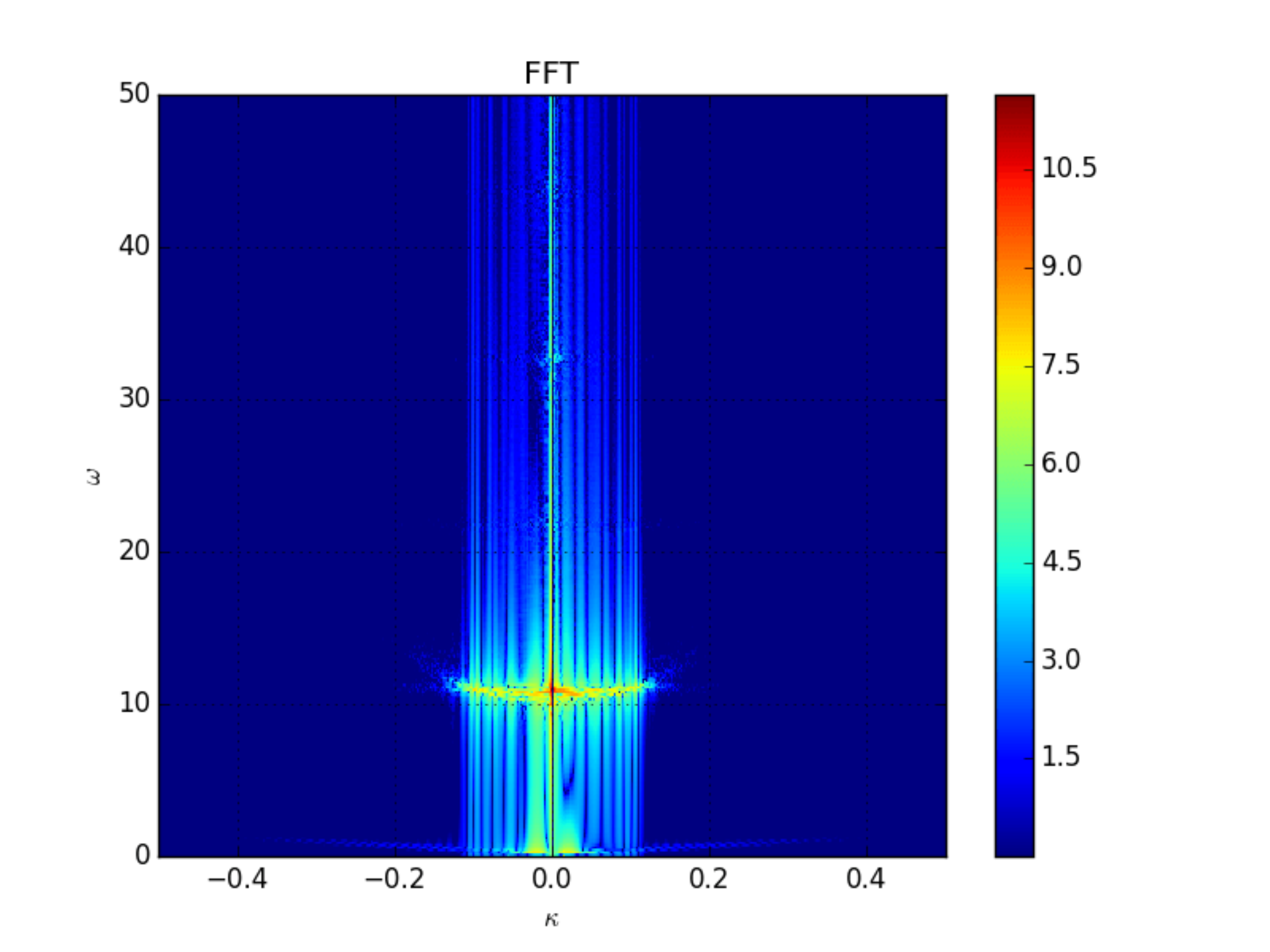}
				   \includegraphics[width=0.25\textwidth]{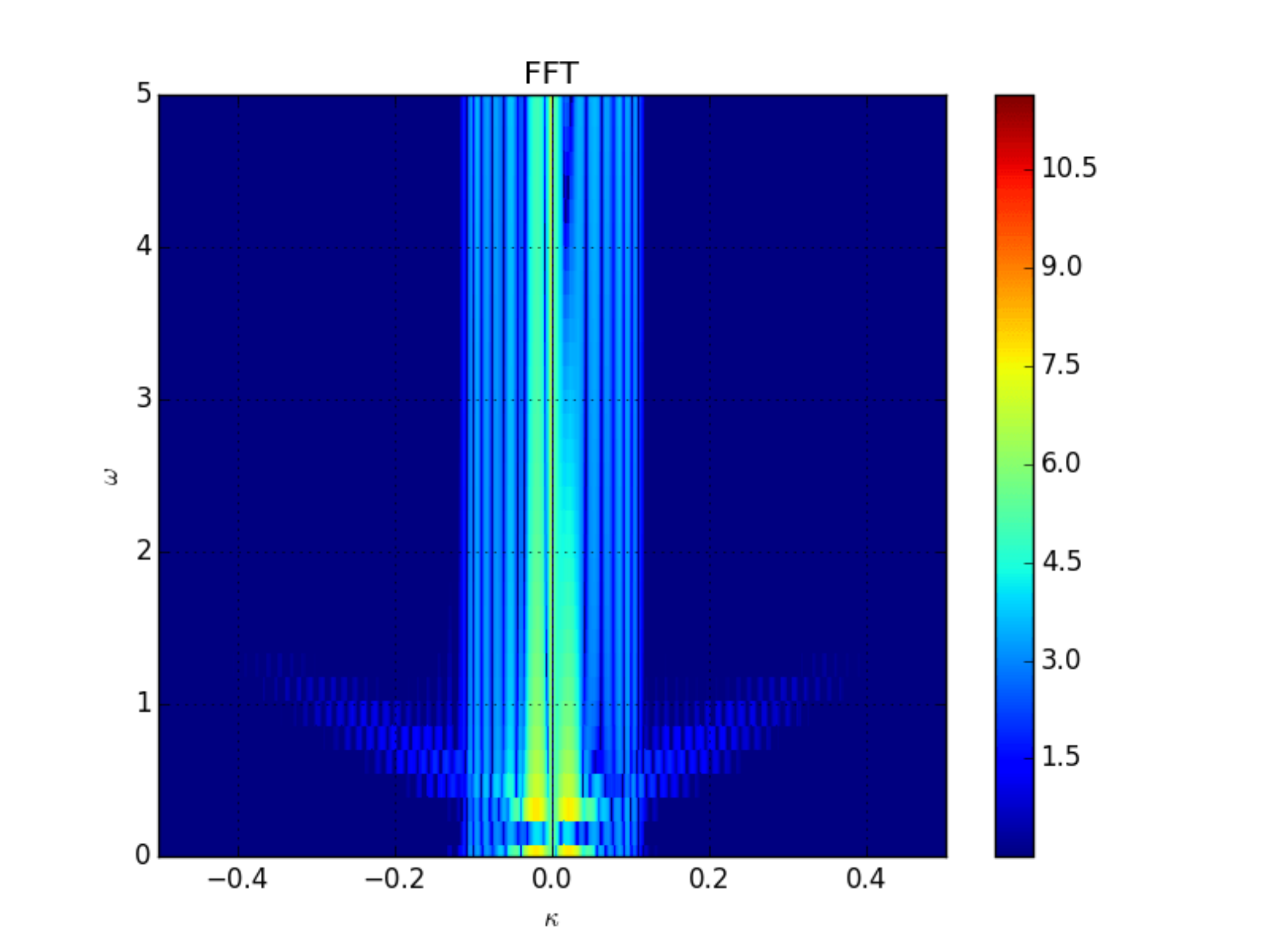}
  \label{T_FFT_2}}
  \caption{The effect of electron-to-ion temperature ratio on the propagation of wavepackets in the disintegration process.
  Fourier transform of electric field for early time of simulations (in the same time span $0 < \tau \leq 40$) is shown in the logarithmic scale for two cases 
  namely $T_e = T_i$ (\ref{T_FFT_1})  and $T_e = 5T_i$ (\ref{T_FFT_2}). 
  The branch of Langmuir waves can be observed in both cases. 
  No ion-acoustic waves can propagate for $T_e = T_i$ ( because of strong Landau damping \cite{jenab2014vlasov}).}
  \label{T_FFT}
\end{figure}

Fig.\ref{T_FFT} represents the two-dimensional Fourier transform of electric field in time for both cases mentioned above. 
Two branches of waves exist in case of $\frac{T_e}{T_i}=5$, namely ion-acoustic and Langmuir waves.
Due to the strong Landau damping, ion-acoustic waves can not propagate in the case of $T_e = T_i $ \cite{jenab2014vlasov}.
\begin{figure}[htp]
   \centering
  \subfloat[$T_e =  T_i$]{\includegraphics[width=0.25\textwidth]{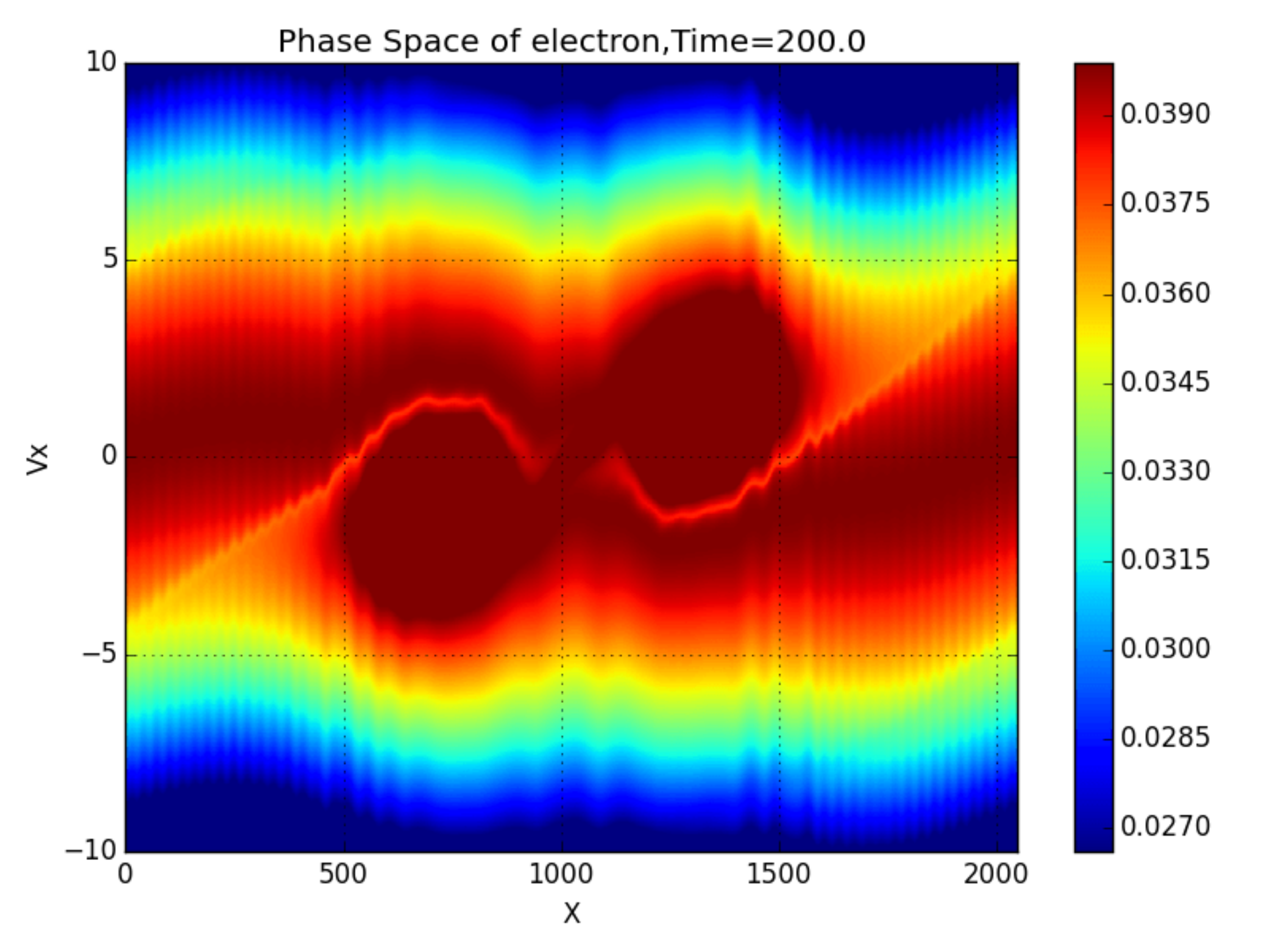}\label{T_DF_1}}
  \subfloat[$T_e =  5 T_i$]{\includegraphics[width=0.25\textwidth]{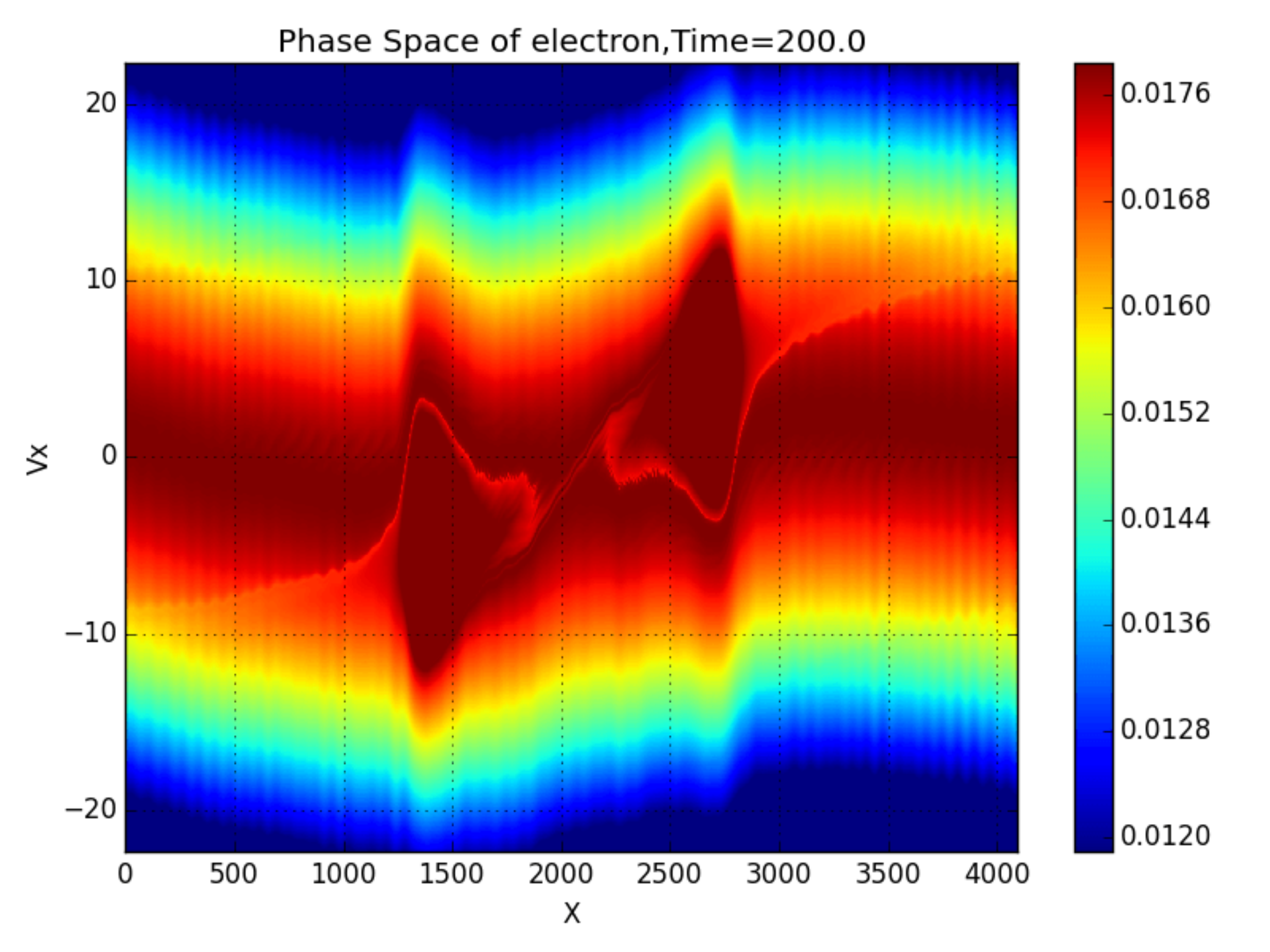}\label{T_DF_2}}
  \caption{The effect of the electron-to-ion temperature ratio on the dynamics of distribution function around moving IDPs.
  The distribution function of electrons for the time ($\tau = 200$) is shown for two cases, 
  namely $T_e = T_i$ (\ref{T_DF_1})  and $T_e = 5T_i$ (\ref{T_DF_2}). 
  In case of $T_e = T_i$, no steepening can be witnessed, 
  while for $T_e = 5T_i$ the steepening effect can be observed for both holes on their propagation sides.}
  \label{T_DF}
\end{figure}
Fig.\ref{T_DF} shows the distribution function of electrons for the two cases considered here.
The reason for the absence of IASWs in the case of $\frac{T_e}{T_i} <3.5$ seems to be the absence of nonlinearity. 
Hence, dispersive effects widen the localized structure without any counter-process of steepening.

\subsection{Effect of Trapped Electrons ($\beta$)} \label{beta}
Table \ref{table} presents the results of the fully kinetic simulations for a wide range of $\beta$,
from $\beta = -0.5$ to $\beta = 10$,
while the amplitudes of moving IDPs and $\theta = 64$ stay the same. 
Four main features of the disintegration process are reported, 
namely the disintegration time ($\tau_d$),
the number ($N_s$), speeds ($v_s$) and sizes ($\delta_x$ and $\delta_v$) 
of the self-consistent IASWs.
The speeds of solitary waves are arranged from
the fastest (which is the most dominant one)
down to the slowest (smallest) one. 
The velocity of each IASWs are measured based on the temporal evolution of number density of the electrons and ions.
The width on spatial and velocity direction are determined based on the symmetry of the electron hole. 
So the width is twice the distance from the center of hole on the associated direction as far as the symmetry exists between two sides of the hole. 
The disintegration time marks the early stage of the appearance of the first soliton as a separate hole in the phase space. 

\begin{table}
\caption{ Simulation results for disintegration process 
and resulting self-consistent IASWs is presented for $\theta = \frac{T_e}{T_i} = 64$.
Features such as number of ion-acoustic solitary waves $N_s$,
their velocities $V_s$, the disintegration times $\tau_d$ 
and their size on x-direction $\delta_x$ and velocity direction $\delta_v$ are reported.
}
\begin{ruledtabular}
\begin{tabular}{cccccccccc}
  \multirow{2}{*}{case}&   \multirow{2}{*}{$\beta$}& 
  \multicolumn{2}{c}{$\tau_d\pm10$}&   \multirow{2}{*}{$N_s$}& 
  \multicolumn{3}{c}{$V_s\pm0.1$}&  \multicolumn{2}{c}{Width} \\
  {}&  {}&  $\tau_{d_1}$&$\tau_{d_2}$&  {}&  $V_{s_1}$&    $V_{s_2}$&  $V_{s_3}$&    $\delta_x$&  $\delta_v$ \\
 \hline
  1     & -0.5  &15&40  & 3  & 12.8&10.2 &8.5    &180$\pm$20  &160$\pm$20  \\
  2     & -0.1  &45&55    & 3  & 10.2&9.4  &8.3  &150$\pm$20  &110$\pm$20\\
  3     & 0     &60&97  & 3  & 9.8 &8.7  &8.2    &140$\pm$20  &105$\pm$20\\
  4     & 0.2   &75&-   & 2  & 9.2 &8.2  &-      &130$\pm$20  &102$\pm$20\\
  5     & 0.5   &90&-   & 2  & 8.5 &7.7  &-      &120$\pm$20  &100$\pm$15 \\
  6     & 1     &80&-   & 2  & 7.7 &6.8  &-      &100$\pm$15  &60$\pm$10\\
  7     & 1.5   &75&-   & 2  & 7.2 &6.7  &-      &80$\pm$10   &50$\pm$10\\
  8     & 3     &70&-   & 2  & 6.1 &5.5  &-      &60$\pm$10   &45$\pm$5\\
  9     & 6     &-&-    & 1  & 4.4 &-  &-        &38$\pm$10   &30$\pm$5\\
  10    & 10    &-&-    & 1  & 4.2 &-    &-      &20$\pm$5    &15$\pm$2\\
\end{tabular}
\end{ruledtabular}
\label{table}
\end{table}

\begin{figure}[htp]
  \centering
  \subfloat[$\beta =0$]{\includegraphics[width=0.25\textwidth]{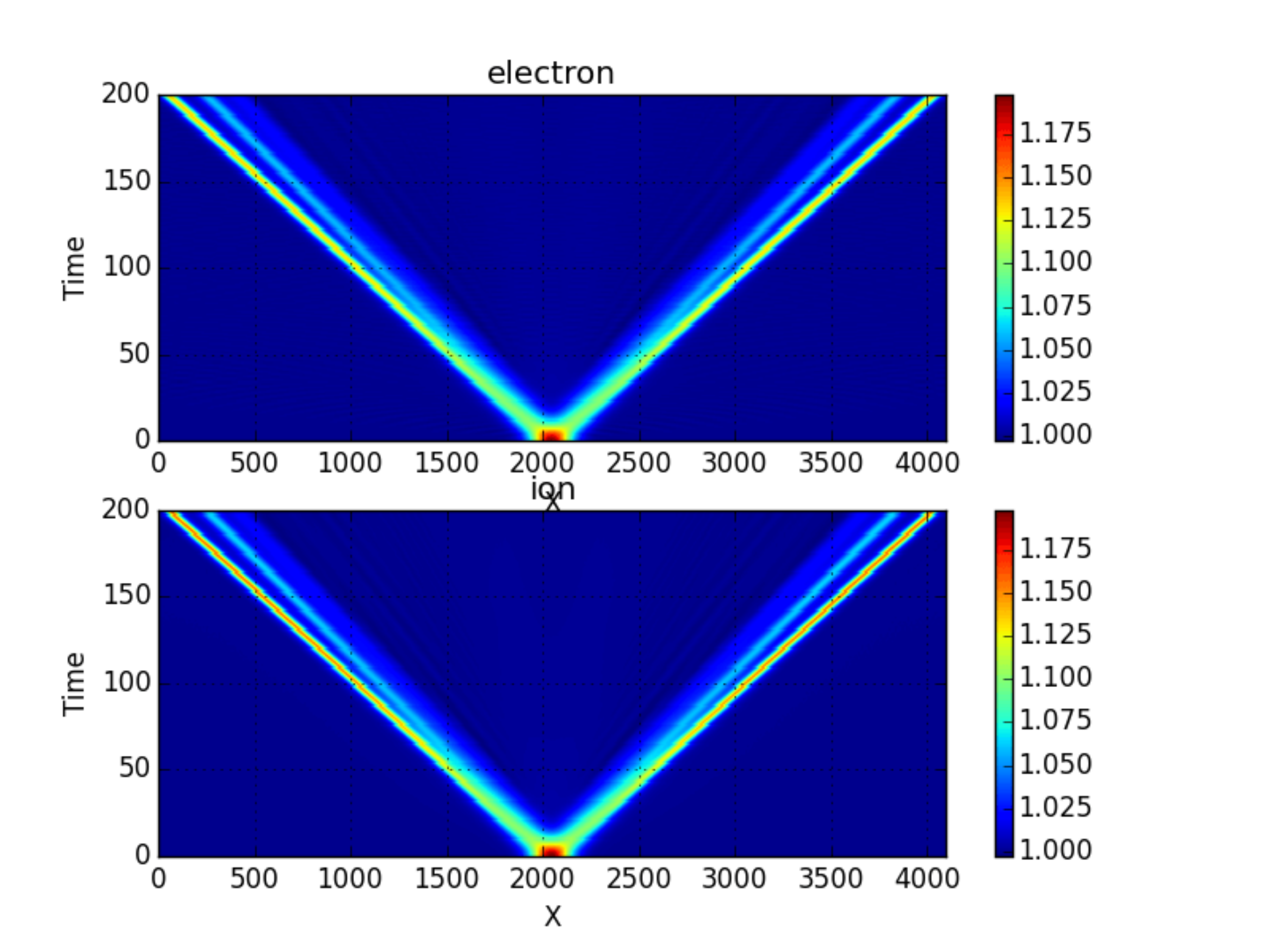}\label{B_0_021_Num_a}} 
  \subfloat[$\beta =0.2$]{\includegraphics[width=0.25\textwidth]{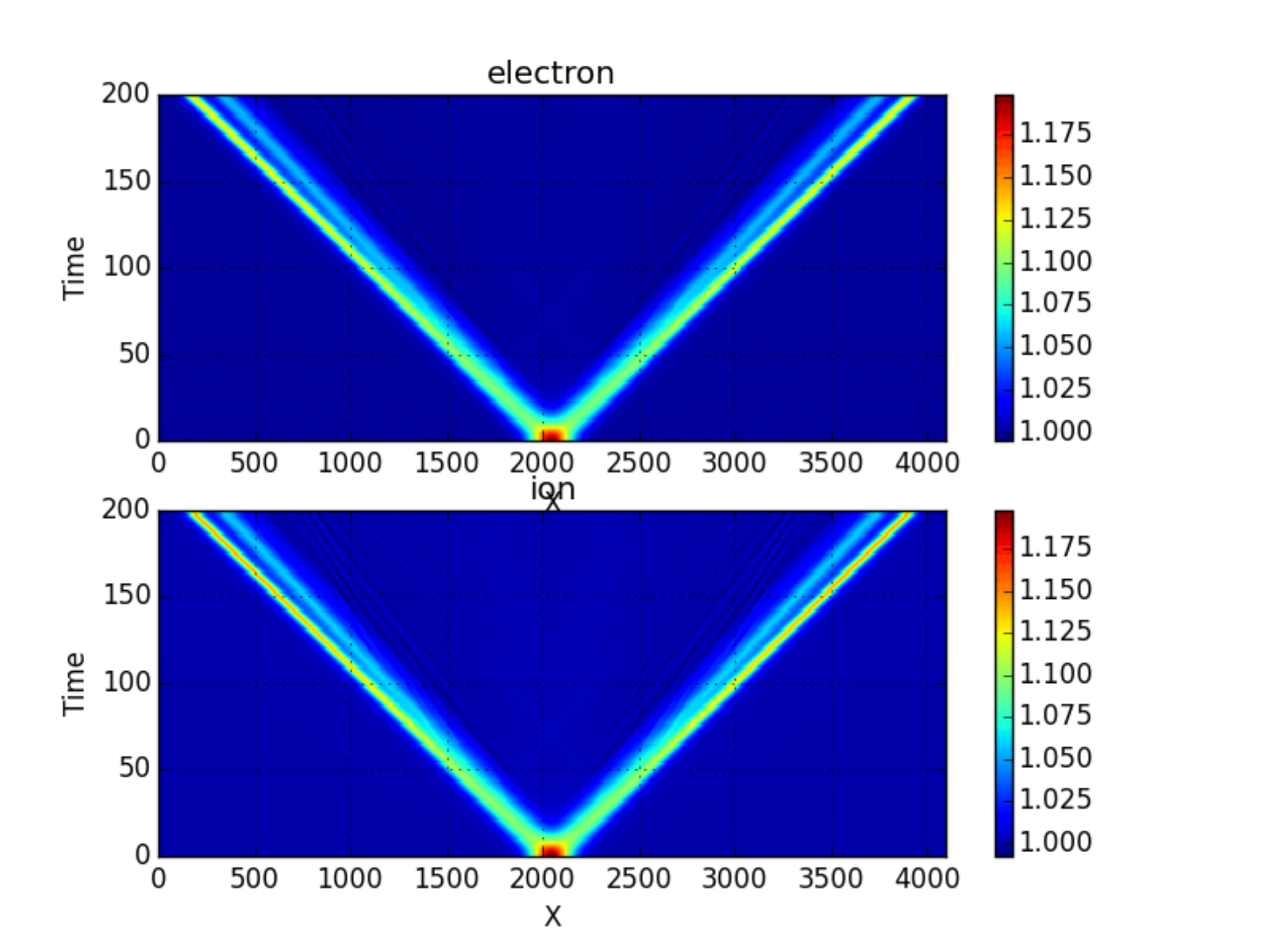}\label{B_0_021_Num_b}}
  \caption{ Symmetrical disintegration/propagation is shown in the temporal evolution of number densities of electrons and ions. 
  In case of $\beta =0$ three solitary waves emerges from any of the moving IDPs (\ref{B_0_021_Num_a}). 
  By increasing trapped electrons $\beta = 0.2$, the number solitary waves drops to two (\ref{B_0_021_Num_b}).}
  \label{B_0_021_Num}
\end{figure}

\begin{figure}[htp]
  \centering
  \begin{tabular}[b]{c}
   \subfloat[number densities ($\beta =0$)]{ \includegraphics[width=0.25\textwidth]{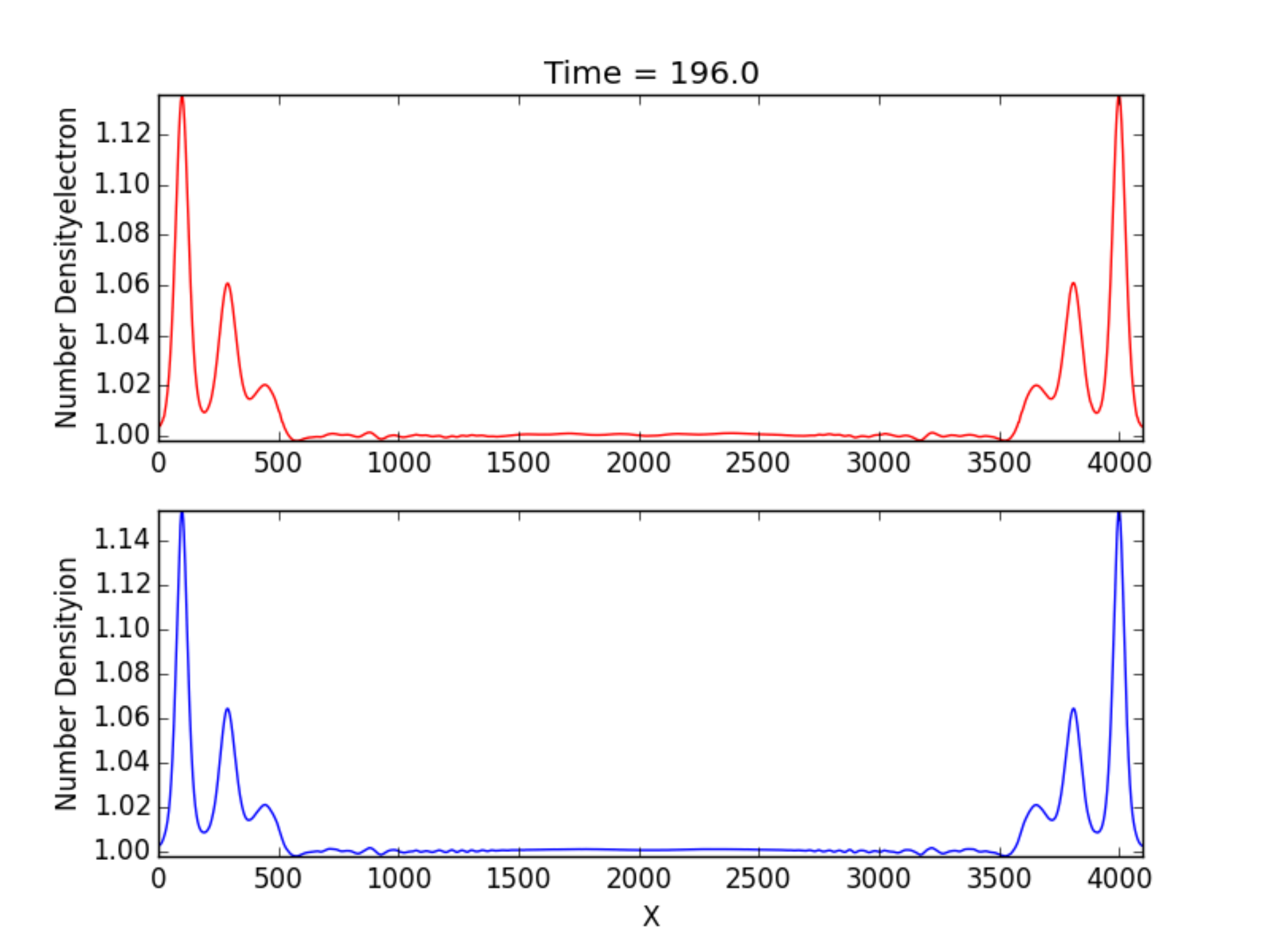}\label{B_0_021_t_a}}
  \subfloat[distribution functions ($\beta =0$)]{  
  \begin{tabular}[b]{c}
        {\includegraphics[width=0.2\textwidth,height=0.1\textwidth]{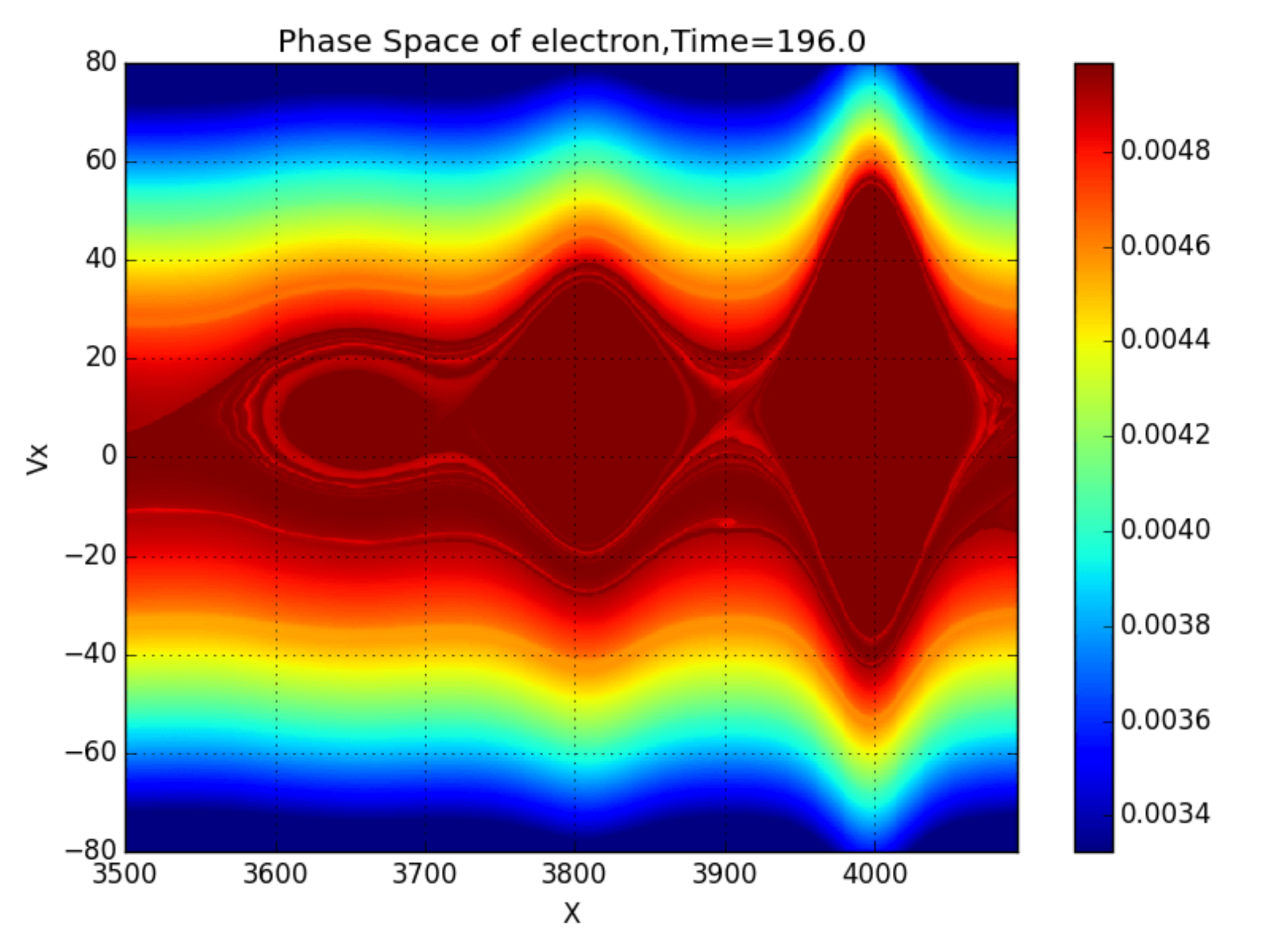}}\\
        {\includegraphics[width=0.2\textwidth,height=0.1\textwidth]{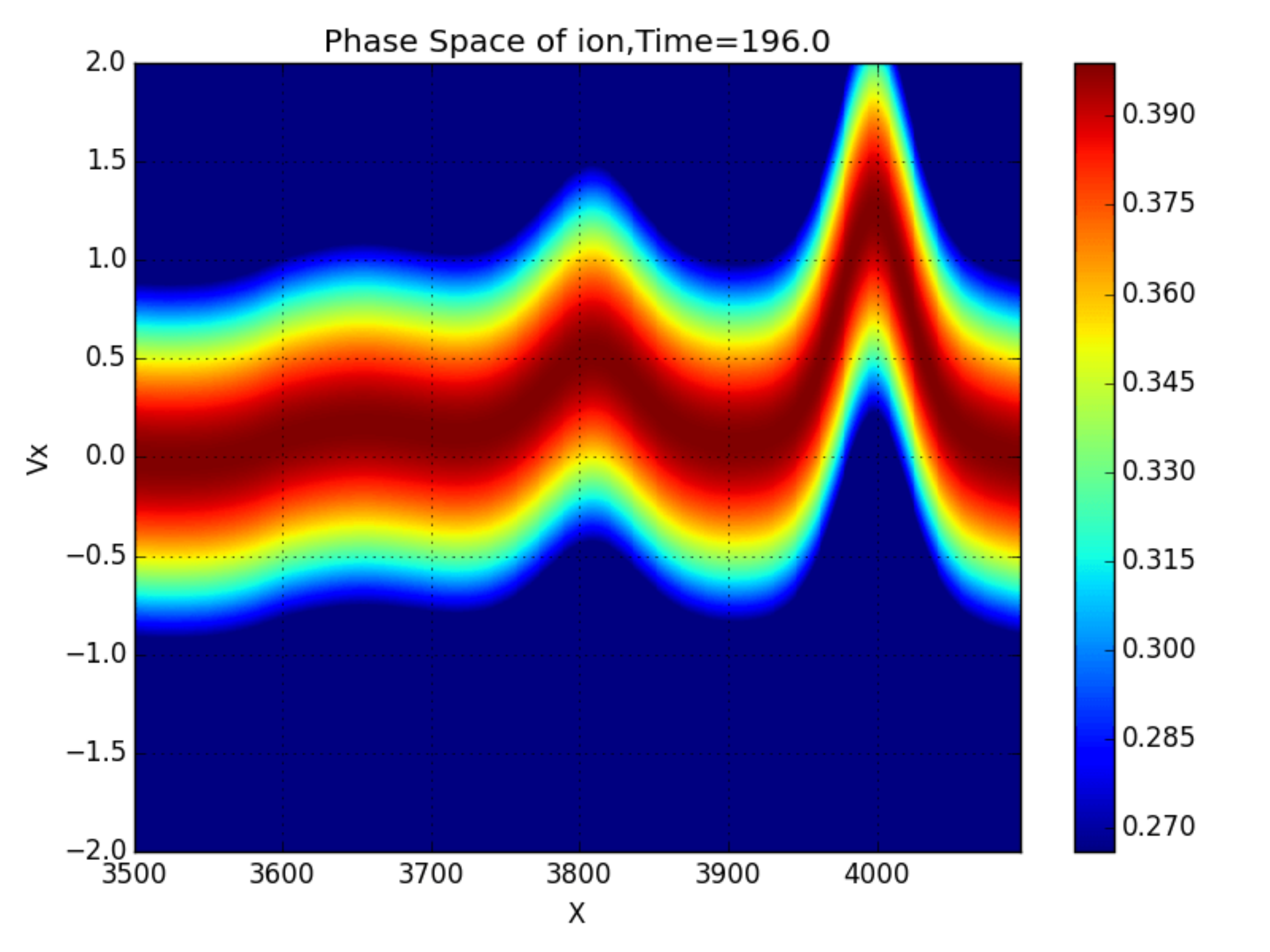}}
  \end{tabular}
  \label{B_0_021_t_b}}\\
  
  \subfloat[number densities ($\beta =0.2$)]{ \includegraphics[width=0.25\textwidth]{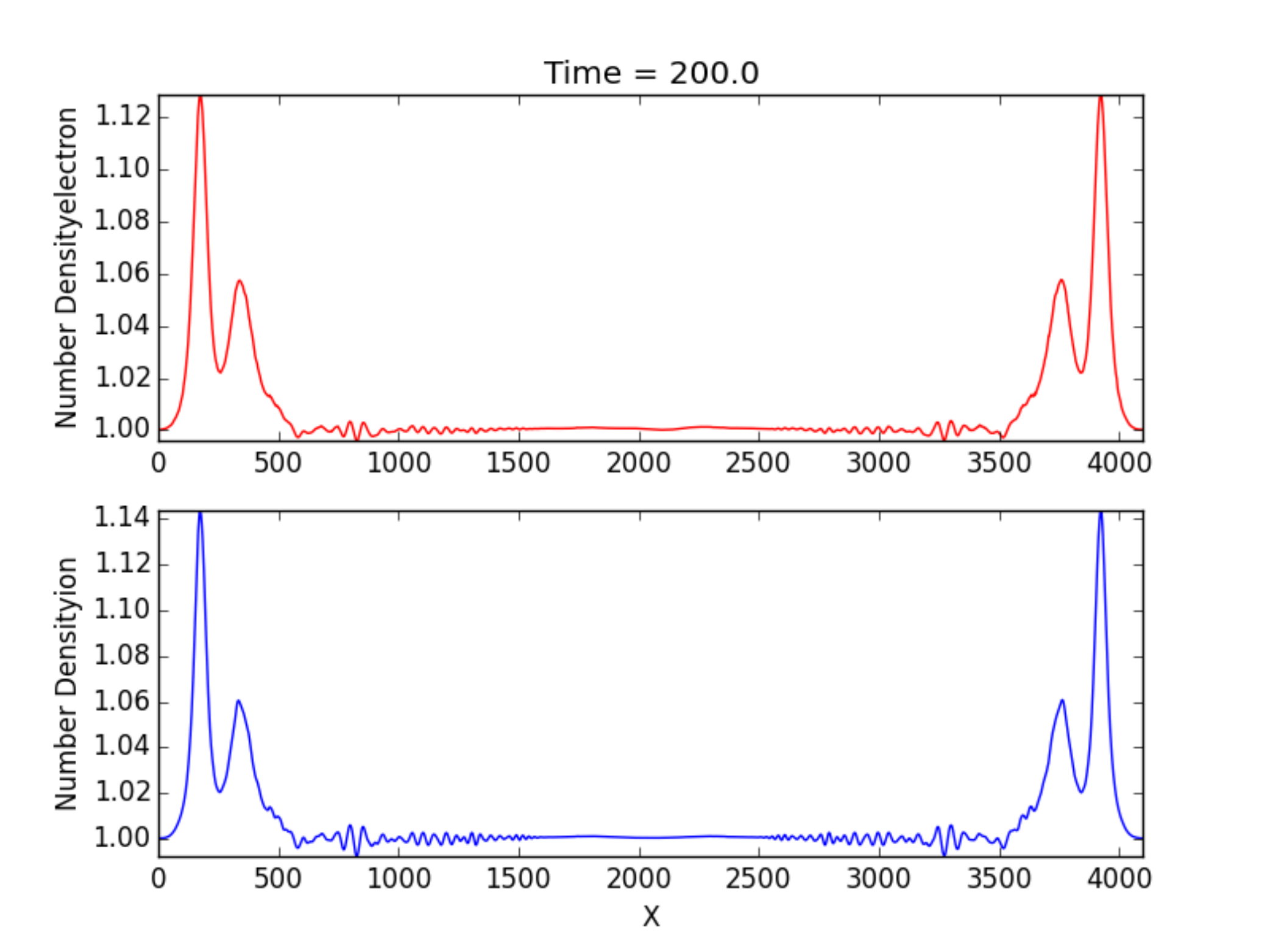}\label{B_0_021_t_c}}
  \subfloat[distribution functions ($\beta =0.2$)]{  
  \begin{tabular}[b]{c}
        \includegraphics[width=0.2\textwidth,height=0.1\textwidth]{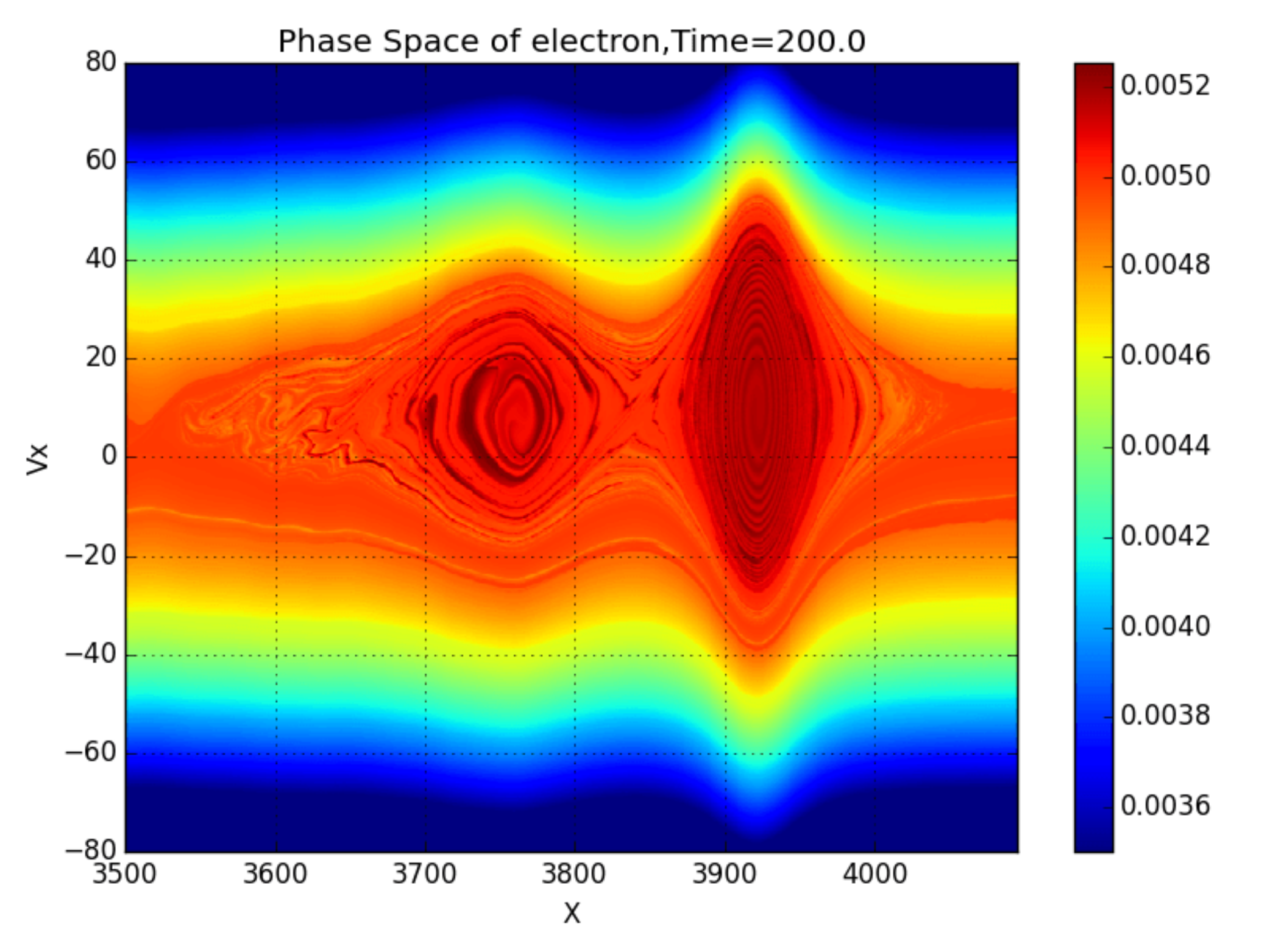} \\
        \includegraphics[width=0.2\textwidth,height=0.1\textwidth]{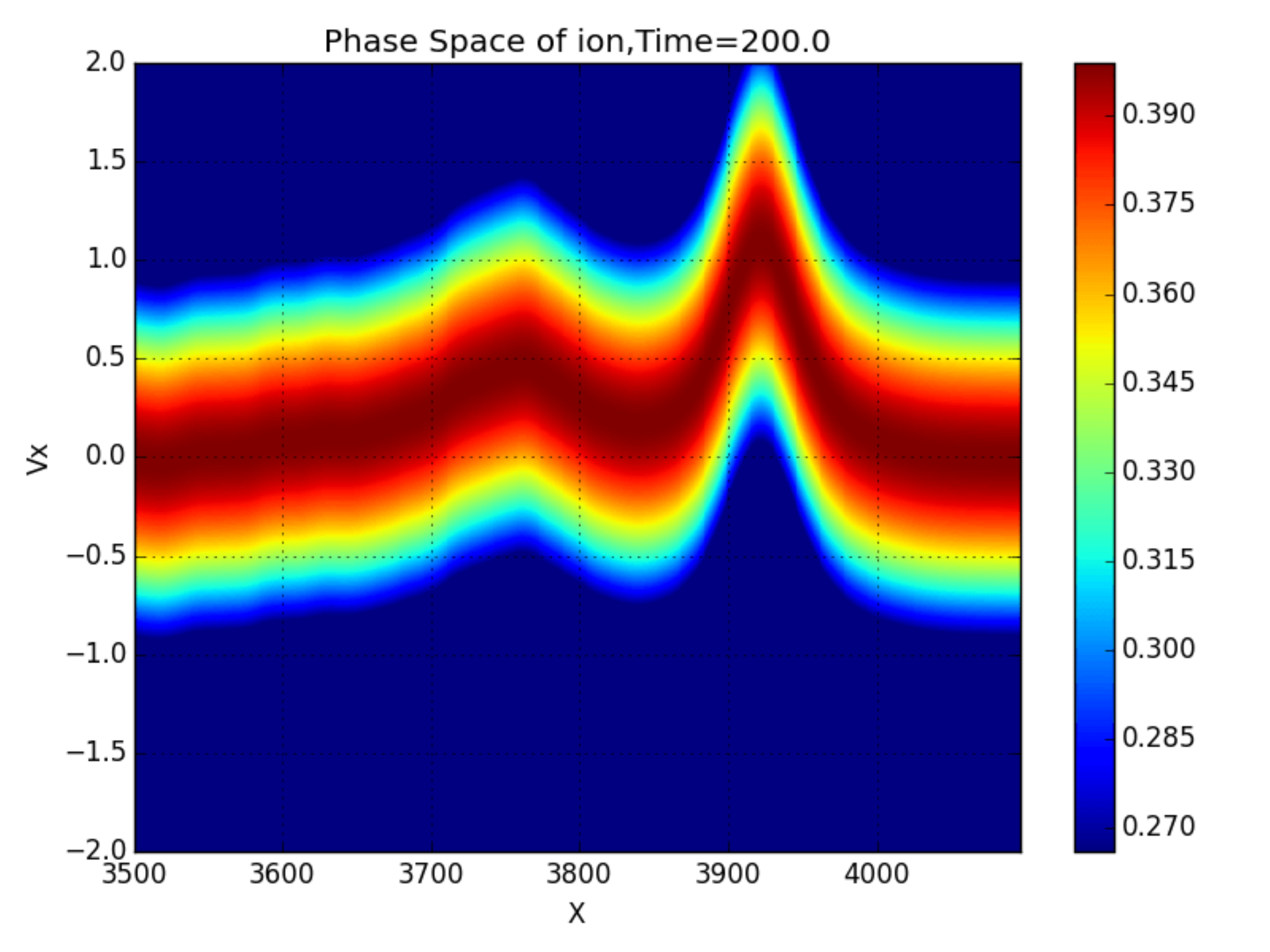}
  \end{tabular}
  \label{B_0_021_t_d}}
  \end{tabular}
  \caption{Solitary waves in two cases $\beta = 0$ with three solitary waves (\ref{B_0_021_t_a},\ref{B_0_021_t_b}),
  and $\beta = 0.2$ with two solitary waves (\ref{B_0_021_t_c},\ref{B_0_021_t_d}) are shown.}
  \label{B_0_021_t}
\end{figure}

Fig.\ref{B_0_021_Num} presents the results of the kinetic simulation 
for the two cases $\beta = 0$ (case 3 in table \ref{table}) 
and $\beta = 0.2$ (case 4 in table \ref{table}),
in which IDPs disintegrate into three and two solitary waves respectively. 
The symmetry in velocity direction during the temporal evolution 
is clear in the figures,
as reported in fluid simulations \cite{Kakad2013}.
The trapping of electrons can be observed 
as a hump and plateau for $\beta = 0.2$ and $\beta =0$ 
in the distribution function respectively (see Fig.\ref{B_0_021_t}). 
On the other hand, no hole, plateau or hump can be witnessed
in the distribution function of ions,
which suggests that the propagation of IASWs in these set of simulations
happens in the absence of the ion trapping.

\begin{figure}[htp]
  \centering
  \subfloat[]{\includegraphics[width=0.38\textwidth]{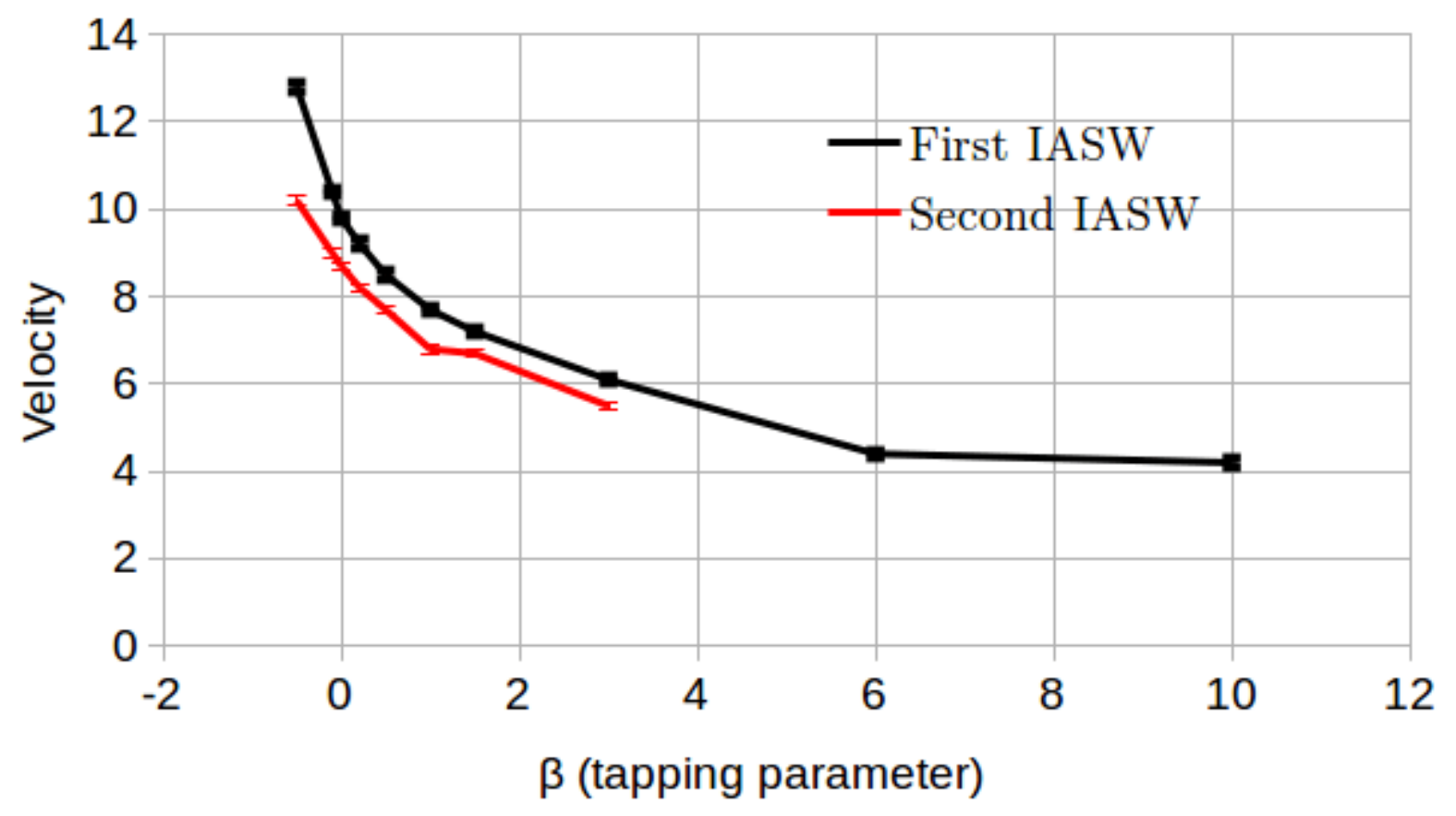}\label{V-B_1}} \\
  \subfloat[]{\includegraphics[width=0.4\textwidth]{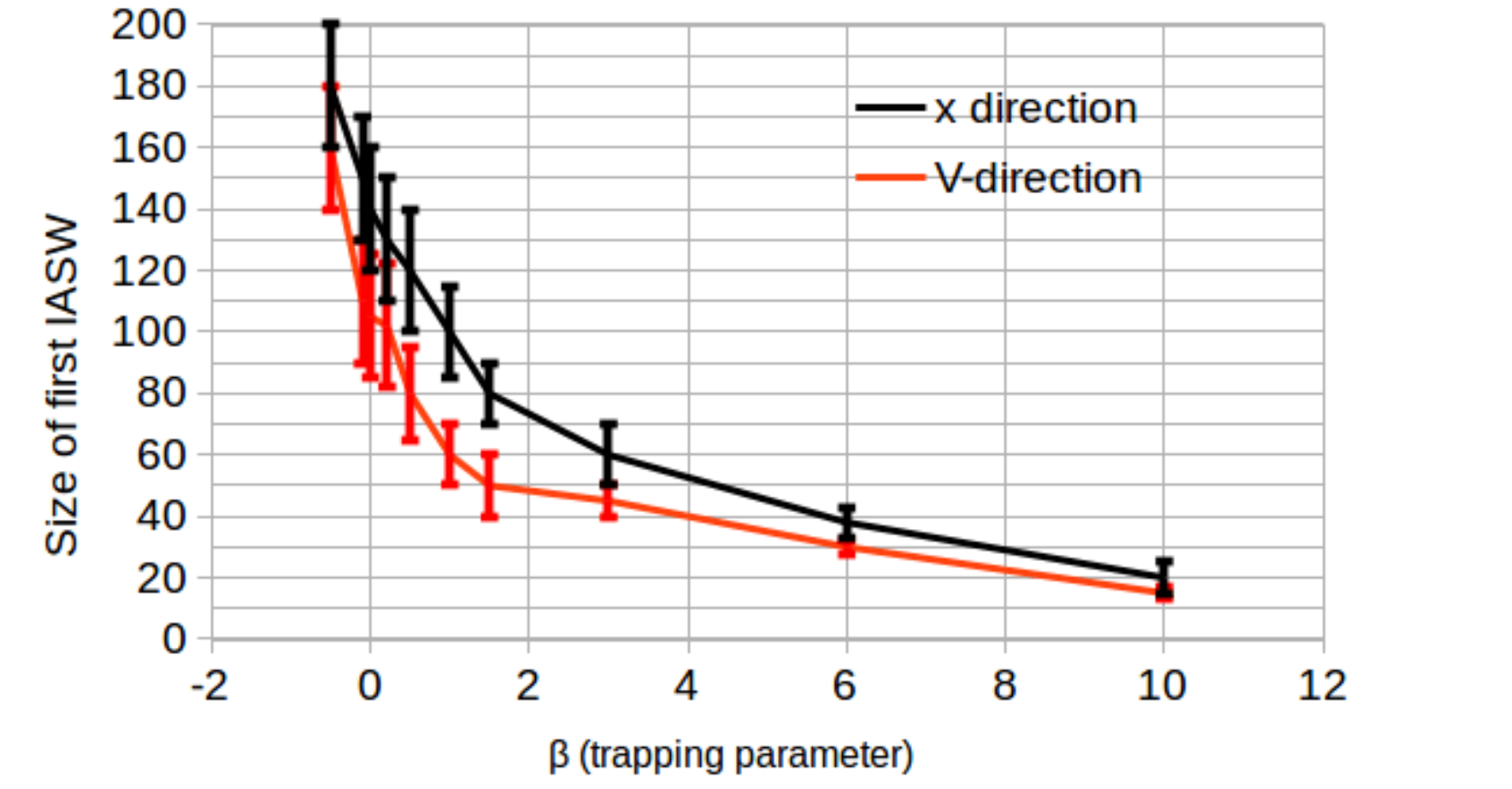}\label{V-B_2}}
  \caption{The dependencies of two different features of IASWs on trapping parameter $\beta$ are presented.
  Velocities of the first and second solitary waves reveals a same decay patterns 
  due to the inertia introduced by higher number of trapped electrons as $\beta$ increases(\ref{V-B_1}).
  The size of the first solitary waves in both spatial and velocity direction (\ref{V-B_2}) 
  drop as $\beta$ rise. Since higher $\beta$ causes stronger nonlinearity and therefore more powerful steepening}
  \label{V-B}
\end{figure}

By increasing $\beta$,
the velocity of the moving IDPs and resulting IASWs decreases,
which is in agreement with the prediction of the nonlinear fluid theory \cite{schamel_3}. 
A hole/hump in the phase space (the trapped electrons) acts a quasi-particle \cite{ghizzo1987bgk} 
and the number of trapped particles is associated with the inertia of such a quasi-particle \cite{schamel_5}. 
Changes in the inertia of the hole/hump would affect its velocity,
and hence propagation velocities of the IASWs.
Therefore,
any increase in $\beta$ results in the decrease of propagation velocities of IASWs. 
As table \ref{table} shows, this tendency appears for each of the three IASWs. 
Fig.\ref{V-B_1} presents the propagation velocities of 
the first and second IASWs versus the trapping parameter $\beta$. 
The same decay behavior can be seen for both solitary waves.
However, the decay saturates to $v=4$, 
which is the half of ion sound in our simulations,
for $\beta \gg 0$.

Concentrating on the first IASW, 
which is also the fastest and the most dominant one in terms of size 
(both in the velocity and spatial direction),
it was observed that any increase in $\beta$ 
decreases its size in both directions (see Fig. \ref{V-B_2}). 
The nonlinear fluid theory
-by associating a nonlinearity to the trapping effect-
suggests that any growth in $\beta$ works in favor of steepening. 
Therefore,
the size of IASW shrinks on x-direction. 
However, kinetic simulation approach shows that
this decline in its spatial size is accompanied 
by the decrease in its size in the velocity direction. 
\begin{figure}[htp]
  \subfloat[$\beta = -0.5$]{\includegraphics[width=0.2\textwidth]{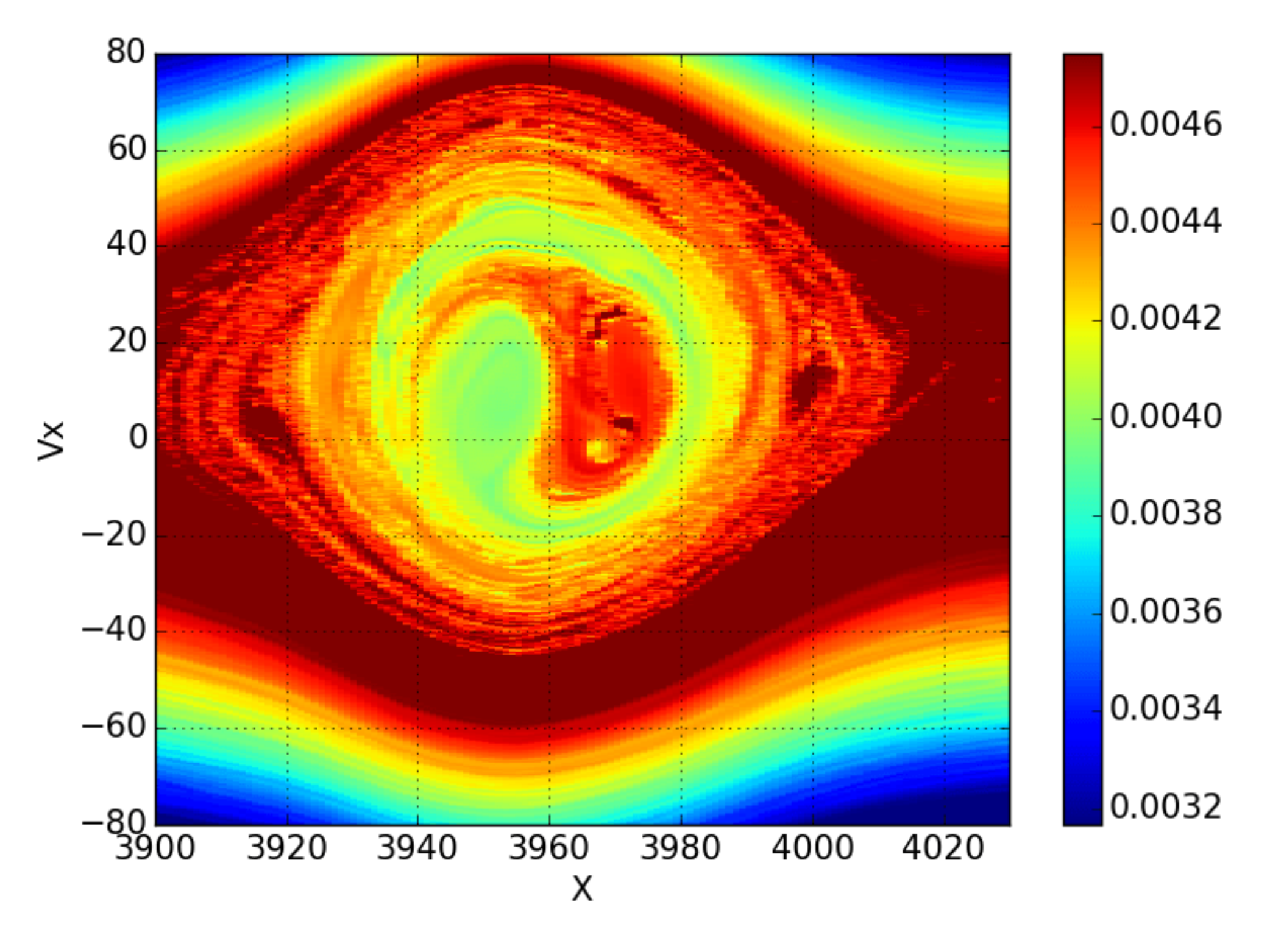}\label{1st_1}}
  \subfloat[$\beta = -0.1$]{\includegraphics[width=0.2\textwidth]{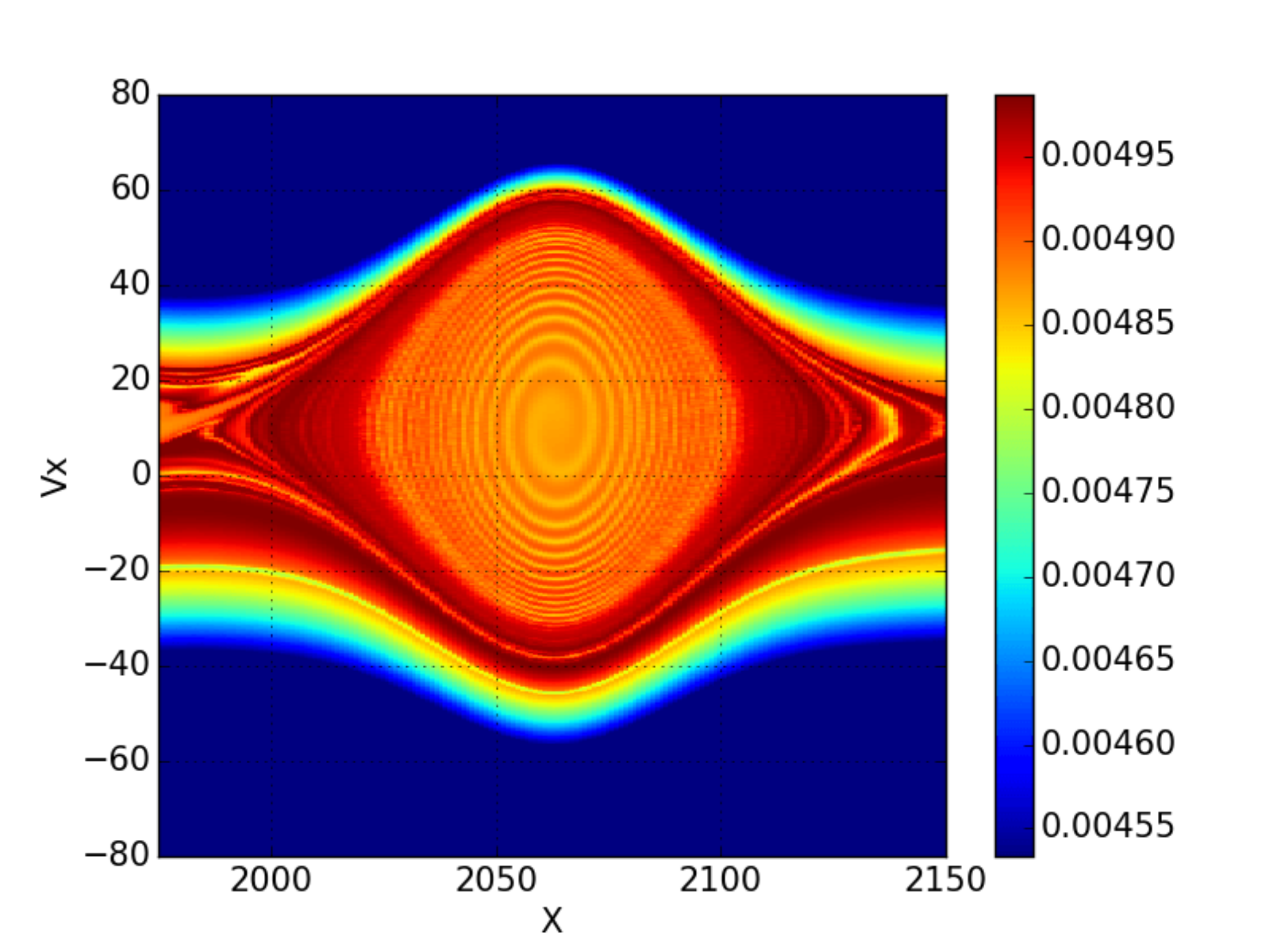}\label{1st_2}}\\
  \subfloat[$\beta = 0$]{\includegraphics[width=0.2\textwidth]{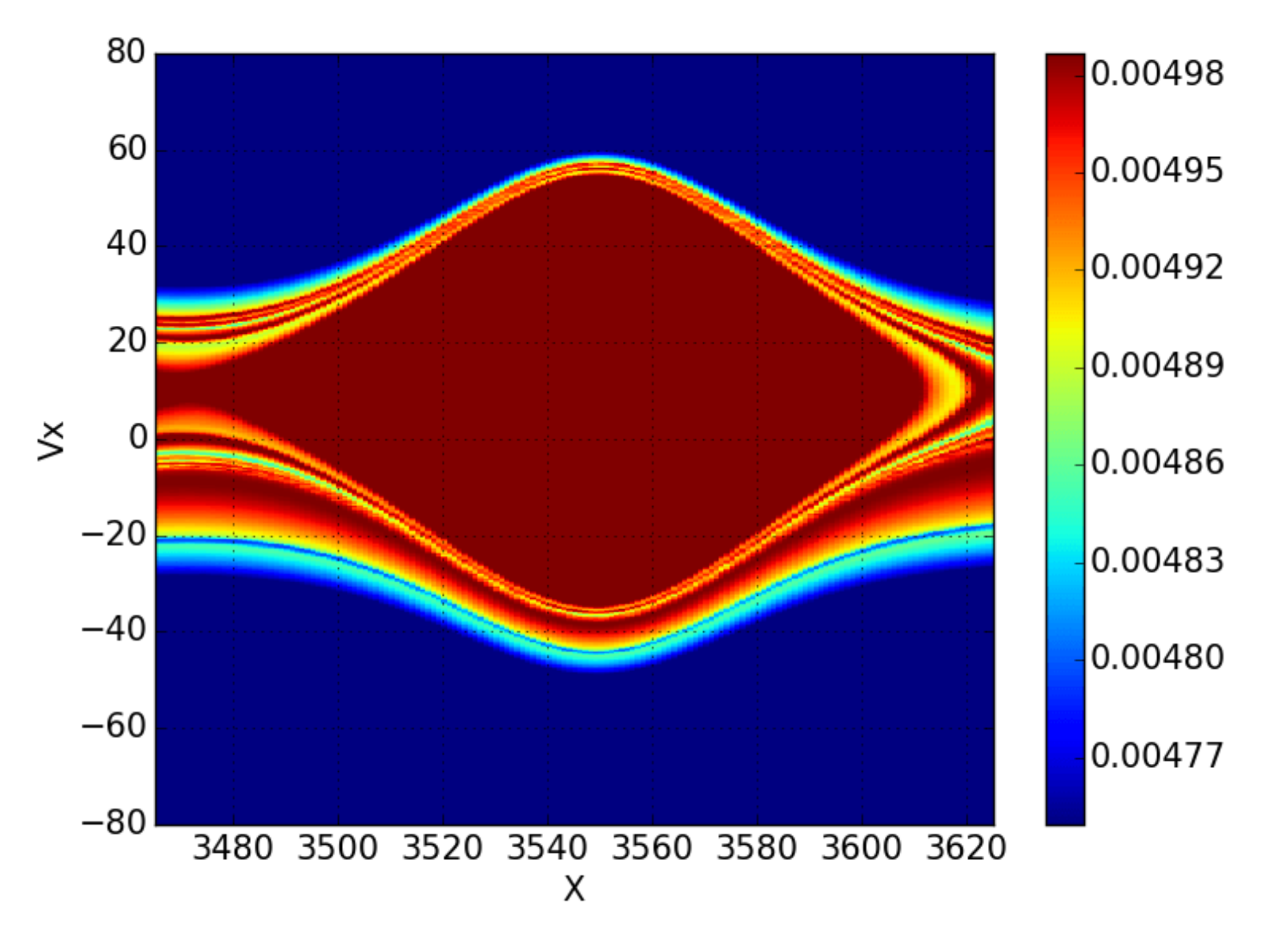}\label{1st_3}}
  \subfloat[$\beta = 0.2$]{\includegraphics[width=0.2\textwidth]{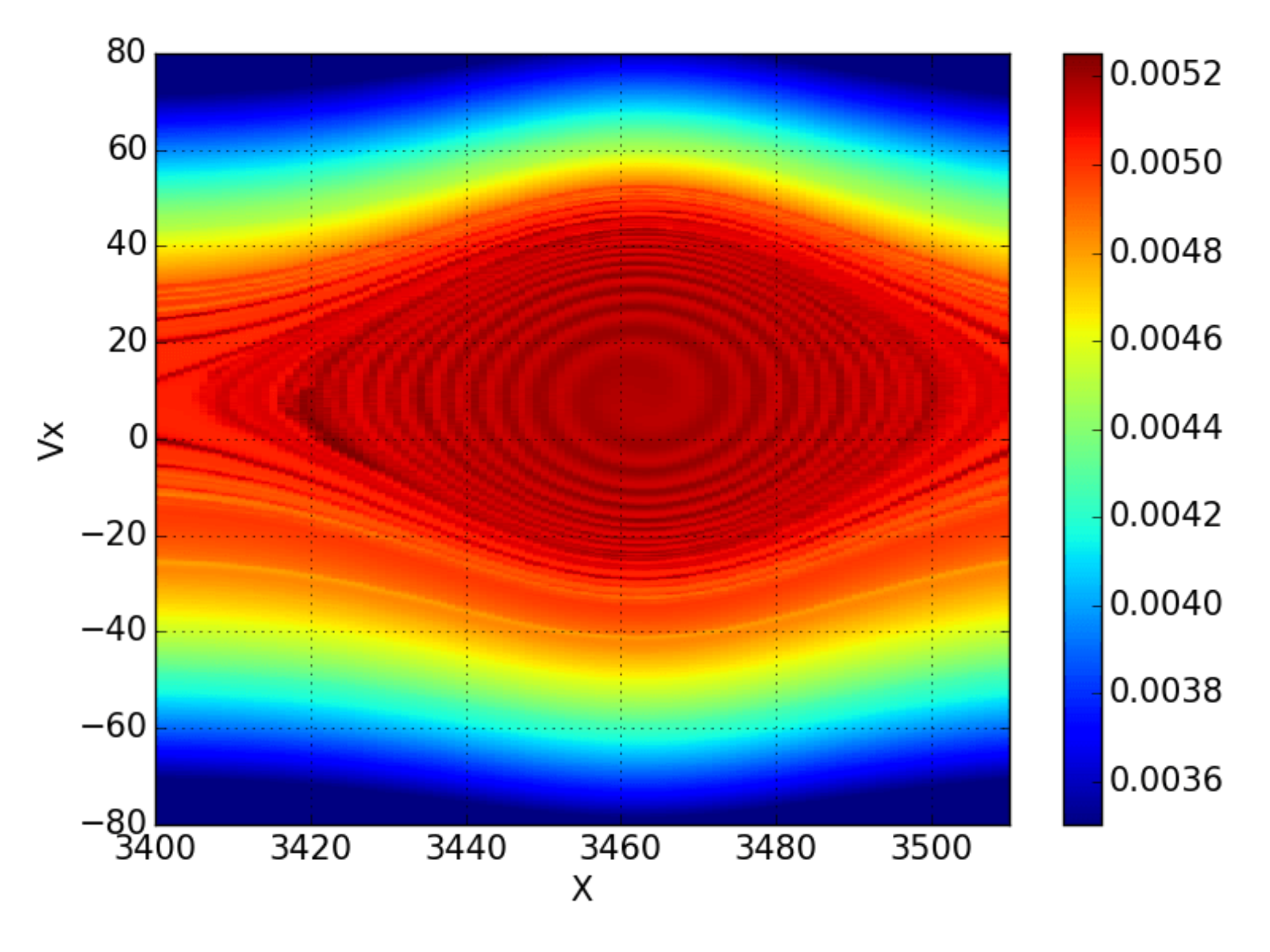}\label{1st_4}}\\
  \subfloat[$\beta = 0.5$]{\includegraphics[width=0.2\textwidth]{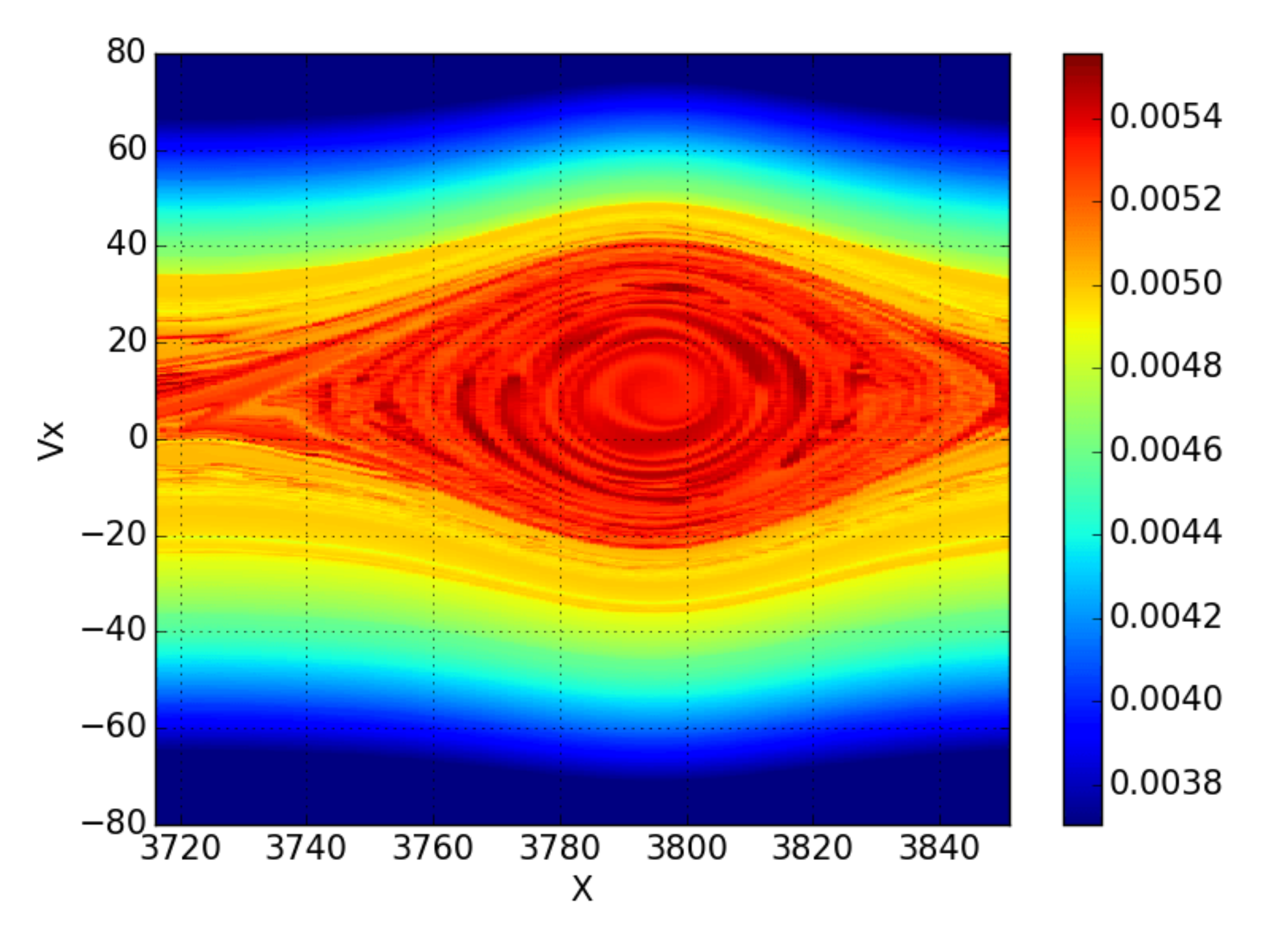}\label{1st_5}}
  \caption{A comparison of nonlinear phase space structures of 
  the dominant/first solitary waves 
  for different value of $\beta$ at time $\tau = 200$ is presented.
  As $\beta \rightarrow 0$, the symmetry of the nonlinear structure grows.}
  \label{1st_soliton_DF}
\end{figure}
It is discussed in the nonlinear fluid theory that for $\beta>1$,
IASWs should be unstable\cite{schamel_3}.
Contrastively, kinetic simulations show the existence and propagation of IASWs
for this range as far as $\beta =10$.

\begin{figure}[htp]
  \centering
  \includegraphics[width=0.4\textwidth]{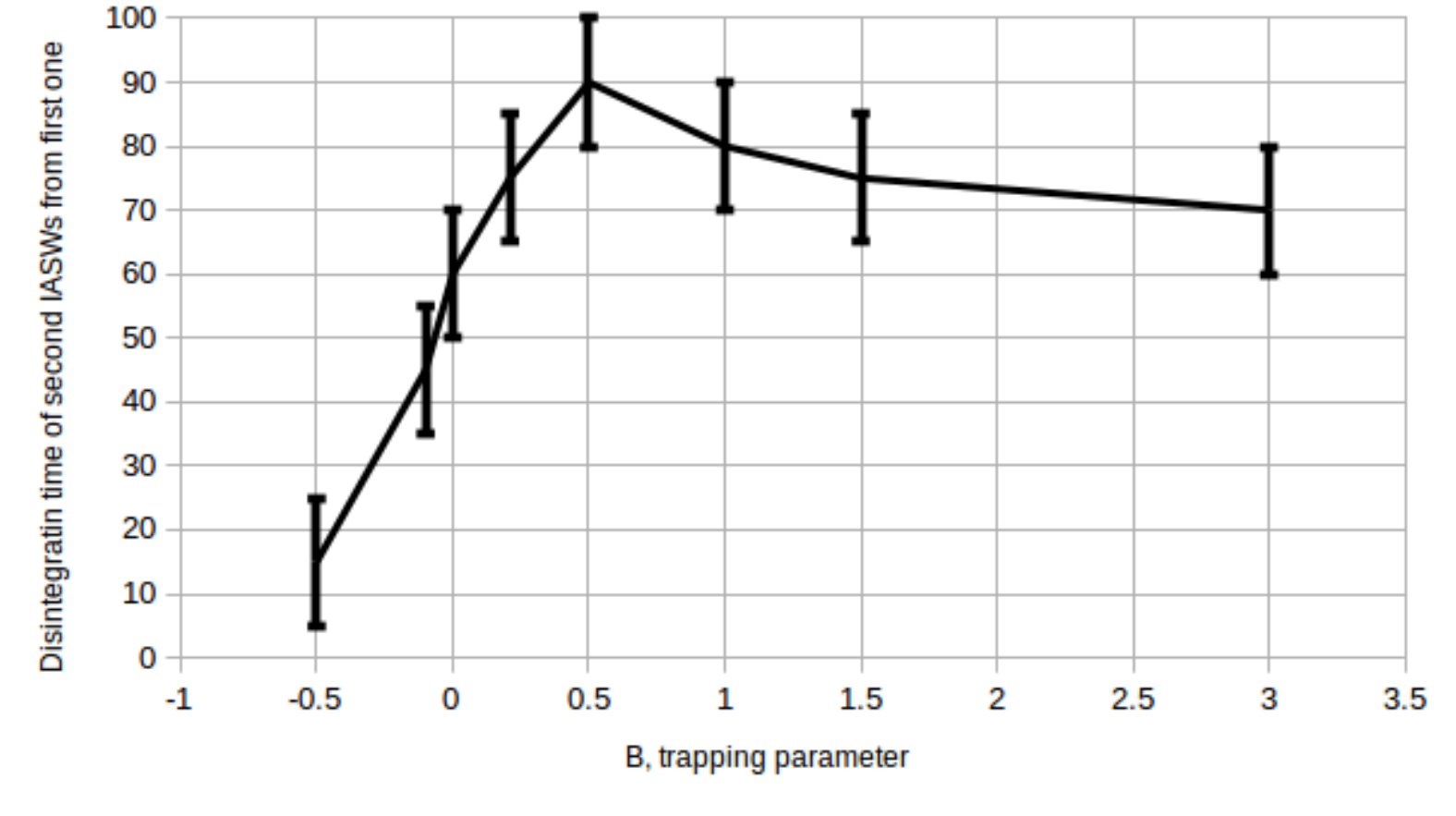}
  \caption{The dependency of the disintegration time on the trapping parameter $\beta$ is shown.
  For $\beta \leq 0$ as $\beta$ increases the disintegration time increases rapidly, 
  while for $\beta >0$, it stays approximately the same. }
  \label{Time-B}
\end{figure}

The disintegration time is defined as the initiation time of the splitting process in the number density graph.
$\tau_{d_1}$ identifies this time for the second IASW, splitting from the first one. 
For the cases of $\beta \leq 0$, a second disintegration time is reported in table \ref{table}, 
which presents the beginning of the splitting process of the third IASW from the second one $\tau_{d_2}$.
The time of disintegration decreases rapidly 
when there is a hole ($\beta \leq 0$) accompanying the IDPs as $\beta$ decreases.
However,
for a hump ($\beta > 0$) accompanying the IDPs,
the time of the disintegration stays more or less the same for a wide range of $\beta>0$ (see Fig.\ref{Time-B}).
The number of IASWs has shown the same tendency as well. 
For the same amplitude of IDPs, the number of IASWs increases
as $\beta$ passes the threshold of $\beta =0$. 
However, for a wide range of $\beta>0$ just two IASWs can be observed.
Note that the disintegration time can not be predicted
by Sagdeev's theoretical method,
since it considers the steady state solutions. 
However, due to the stronger nonlinearity $|b|\geq \sqrt{\psi}$ arising from the trapped electrons,
a greater number of solitary waves 
should appear on a shorter timescale \cite{schamel_4}.
Kinetic simulation results,
contrastively,
imply that a growth in $\beta$ (for $\beta >0$) 
does not associate with an increase in the number of IASWs
or a decrease in the disintegration time.
Furthermore, as $\beta$ closes in to zero,
the structure of the accompanying hole/hump 
becomes more symmetric. 
Fig.\ref{1st_soliton_DF} represents the phase space structure
for different values of $\beta$. 

\section{Conclusions} \label{Conclusions}
The trapping effect of electrons on the disintegration of 
an initial density perturbation (IDP) into ion-acoustic solitary waves (IASWs) and the temporal evolution of
these waves have been studied in a wide range of $\beta$. 
Four main features of these dynamics (reported in table \ref{table}) 
and their dependency on $\beta$ have been focused upon (see Figs.\ref{V-B},\ref{Time-B}). 
These dependencies show a smooth and well-defined behavior bridging among all the three theoretical regimes.
These regimes in the nonlinear fluid theoretical approach are considered separately, 
and their solitary wave solutions include different shapes and forms.
The smooth transition among these regimes suggests 
that there should be a general theoretical frame containing all the three regimes. 

Qualitatively speaking,
the nonlinear fluid theory approach suggests some predictions 
about the effect of trapped electrons on the different features 
studied here.
These predictions are validated through fully kinetic simulation approach in this study. 
The rise in the number of trapped electrons (increase in $\beta$)
should increase the inertia of the hole/hump accompanying the solitary waves. 
This can be traced and confirmed in a decay of the propagation velocity of solitary waves.
It is shown here that such a dependency 
follows an exponential decrease,
which is presented in table \ref{table}. 
This specific dependency (to the best of our knowledge) is yet to be reported in the nonlinear theoretical approach. 
On the other hand,
any increase in $\beta$ should increase the (trapping) nonlinearity which results in stronger steepening. 
This would change the balance 
between the steepening and widening tendency inside the dynamics of solitary waves
in favor of steepening.
Therefore, the size of the solitary wave should be reduced due to the stronger steepening \cite{schamel_4}. 
The fully kinetic simulation approach confirms such dependency,
and furthermore shows that 
it follows an exponential decay as $\beta$ increases.
The same tendency can also be witnessed on velocity direction.
In terms of the time of disintegration,
the theoretical suggestion based on the nonlinear fluid theory,
predicts a drop in this time,
while the simulation results here shows a more complicated dependency (see Fig. \ref{Time-B}). 
The same discrepancy occurs in terms of the  number of solitary waves. 
For $\beta$ around zero, the hole or hump takes a symmetric structure (here is shown for $\beta =-0.1$, $0.2$).
As $|\beta|$ grows, the symmetry of the nonlinear structure gets distorted.

\acknowledgments
M. Jenab is thankful to Prof. H. Schamel for his insightful comments.
This work is based upon research supported by the National Research Foundation 
and Department of Science and Technology.
Any opinion, findings and conclusions or recommendations expressed in this 
material are those of the authors and therefore the NRF and DST do not accept 
any liability in regard thereto.


\begin{thebibliography}{45}%
\makeatletter
\providecommand \@ifxundefined [1]{%
 \@ifx{#1\undefined}
}%
\providecommand \@ifnum [1]{%
 \ifnum #1\expandafter \@firstoftwo
 \else \expandafter \@secondoftwo
 \fi
}%
\providecommand \@ifx [1]{%
 \ifx #1\expandafter \@firstoftwo
 \else \expandafter \@secondoftwo
 \fi
}%
\providecommand \natexlab [1]{#1}%
\providecommand \enquote  [1]{``#1''}%
\providecommand \bibnamefont  [1]{#1}%
\providecommand \bibfnamefont [1]{#1}%
\providecommand \citenamefont [1]{#1}%
\providecommand \href@noop [0]{\@secondoftwo}%
\providecommand \href [0]{\begingroup \@sanitize@url \@href}%
\providecommand \@href[1]{\@@startlink{#1}\@@href}%
\providecommand \@@href[1]{\endgroup#1\@@endlink}%
\providecommand \@sanitize@url [0]{\catcode `\\12\catcode `\$12\catcode
  `\&12\catcode `\#12\catcode `\^12\catcode `\_12\catcode `\%12\relax}%
\providecommand \@@startlink[1]{}%
\providecommand \@@endlink[0]{}%
\providecommand \url  [0]{\begingroup\@sanitize@url \@url }%
\providecommand \@url [1]{\endgroup\@href {#1}{\urlprefix }}%
\providecommand \urlprefix  [0]{URL }%
\providecommand \Eprint [0]{\href }%
\providecommand \doibase [0]{http://dx.doi.org/}%
\providecommand \selectlanguage [0]{\@gobble}%
\providecommand \bibinfo  [0]{\@secondoftwo}%
\providecommand \bibfield  [0]{\@secondoftwo}%
\providecommand \translation [1]{[#1]}%
\providecommand \BibitemOpen [0]{}%
\providecommand \bibitemStop [0]{}%
\providecommand \bibitemNoStop [0]{.\EOS\space}%
\providecommand \EOS [0]{\spacefactor3000\relax}%
\providecommand \BibitemShut  [1]{\csname bibitem#1\endcsname}%
\let\auto@bib@innerbib\@empty
\bibitem [{\citenamefont {Abbasi}, \citenamefont {Jenab},\ and\ \citenamefont
  {Pajouh}(2011)}]{jenab2011preventing}%
  \BibitemOpen
  \bibfield  {author} {\bibinfo {author} {\bibnamefont {Abbasi}, \bibfnamefont
  {H.}}, \bibinfo {author} {\bibnamefont {Jenab}, \bibfnamefont {M.}}, \ and\
  \bibinfo {author} {\bibnamefont {Pajouh}, \bibfnamefont {H.~H.}},\
  }\href@noop {} {\bibfield  {journal} {\bibinfo  {journal} {Physical Review
  E}\ }\textbf {\bibinfo {volume} {84}},\ \bibinfo {pages} {036702} (\bibinfo
  {year} {2011})}\BibitemShut {NoStop}%
\bibitem [{\citenamefont {Baluku}, \citenamefont {Hellberg},\ and\
  \citenamefont {Verheest}(2010)}]{baluku2010new}%
  \BibitemOpen
  \bibfield  {author} {\bibinfo {author} {\bibnamefont {Baluku}, \bibfnamefont
  {T.~K.}}, \bibinfo {author} {\bibnamefont {Hellberg}, \bibfnamefont {M.~A.}},
  \ and\ \bibinfo {author} {\bibnamefont {Verheest}, \bibfnamefont {F.}},\
  }\href@noop {} {\bibfield  {journal} {\bibinfo  {journal} {EPL (Europhysics
  Letters)}\ }\textbf {\bibinfo {volume} {91}},\ \bibinfo {pages} {15001}
  (\bibinfo {year} {2010})}\BibitemShut {NoStop}%
\bibitem [{\citenamefont {Bharuthram}\ and\ \citenamefont
  {Shukla}(1992)}]{bharuthram1992large}%
  \BibitemOpen
  \bibfield  {author} {\bibinfo {author} {\bibnamefont {Bharuthram},
  \bibfnamefont {R.}}\ and\ \bibinfo {author} {\bibnamefont {Shukla},
  \bibfnamefont {P.}},\ }\href@noop {} {\bibfield  {journal} {\bibinfo
  {journal} {Planetary and space science}\ }\textbf {\bibinfo {volume} {40}},\
  \bibinfo {pages} {973} (\bibinfo {year} {1992})}\BibitemShut {NoStop}%
\bibitem [{\citenamefont {Catte115}\ \emph {et~al.}(1998)\citenamefont
  {Catte115}, \citenamefont {Klumparfi}, \citenamefont {Shelley}, \citenamefont
  {Petersonfi}, \citenamefont {Moebius},\ and\ \citenamefont
  {Kistler}}]{catte1151998fast}%
  \BibitemOpen
  \bibfield  {author} {\bibinfo {author} {\bibnamefont {Catte115},
  \bibfnamefont {C.}}, \bibinfo {author} {\bibnamefont {Klumparfi},
  \bibfnamefont {D.}}, \bibinfo {author} {\bibnamefont {Shelley}, \bibfnamefont
  {E.}}, \bibinfo {author} {\bibnamefont {Petersonfi}, \bibfnamefont {W.}},
  \bibinfo {author} {\bibnamefont {Moebius}, \bibfnamefont {E.}}, \ and\
  \bibinfo {author} {\bibnamefont {Kistler}, \bibfnamefont {L.}},\ }\href@noop
  {} {\bibfield  {journal} {\bibinfo  {journal} {Geophysical Research Letters}\
  }\textbf {\bibinfo {volume} {25}},\ \bibinfo {pages} {2041} (\bibinfo {year}
  {1998})}\BibitemShut {NoStop}%
\bibitem [{\citenamefont {Cooney}, \citenamefont {Gavin},\ and\ \citenamefont
  {Lonngren}(1991)}]{cooney1991experiments}%
  \BibitemOpen
  \bibfield  {author} {\bibinfo {author} {\bibnamefont {Cooney}, \bibfnamefont
  {J.~L.}}, \bibinfo {author} {\bibnamefont {Gavin}, \bibfnamefont {M.~T.}}, \
  and\ \bibinfo {author} {\bibnamefont {Lonngren}, \bibfnamefont {K.~E.}},\
  }\href@noop {} {\bibfield  {journal} {\bibinfo  {journal} {Physics of Fluids
  B: Plasma Physics (1989-1993)}\ }\textbf {\bibinfo {volume} {3}},\ \bibinfo
  {pages} {2758} (\bibinfo {year} {1991})}\BibitemShut {NoStop}%
\bibitem [{\citenamefont {Debnath}(2007)}]{Debnath20071003}%
  \BibitemOpen
  \bibfield  {author} {\bibinfo {author} {\bibnamefont {Debnath}, \bibfnamefont
  {L.}},\ }\href@noop {} {\bibfield  {journal} {\bibinfo  {journal}
  {International Journal of Mathematical Education in Science and Technology}\
  }\textbf {\bibinfo {volume} {38}},\ \bibinfo {pages} {1003} (\bibinfo {year}
  {2007})}\BibitemShut {NoStop}%
\bibitem [{\citenamefont {Franz}, \citenamefont {Kintner},\ and\ \citenamefont
  {Pickett}(1998)}]{franz1998polar}%
  \BibitemOpen
  \bibfield  {author} {\bibinfo {author} {\bibnamefont {Franz}, \bibfnamefont
  {J.~R.}}, \bibinfo {author} {\bibnamefont {Kintner}, \bibfnamefont {P.~M.}},
  \ and\ \bibinfo {author} {\bibnamefont {Pickett}, \bibfnamefont {J.~S.}},\
  }\href@noop {} {\bibfield  {journal} {\bibinfo  {journal} {Geophysical
  research letters}\ }\textbf {\bibinfo {volume} {25}},\ \bibinfo {pages}
  {1277} (\bibinfo {year} {1998})}\BibitemShut {NoStop}%
\bibitem [{\citenamefont {Gardner}\ \emph {et~al.}(1967)\citenamefont
  {Gardner}, \citenamefont {Greene}, \citenamefont {Kruskal},\ and\
  \citenamefont {Miura}}]{Gardner19671095}%
  \BibitemOpen
  \bibfield  {author} {\bibinfo {author} {\bibnamefont {Gardner}, \bibfnamefont
  {C.~S.}}, \bibinfo {author} {\bibnamefont {Greene}, \bibfnamefont {J.~M.}},
  \bibinfo {author} {\bibnamefont {Kruskal}, \bibfnamefont {M.~D.}}, \ and\
  \bibinfo {author} {\bibnamefont {Miura}, \bibfnamefont {R.~M.}},\ }\href@noop
  {} {\bibfield  {journal} {\bibinfo  {journal} {Physical Review Letters}\
  }\textbf {\bibinfo {volume} {19}},\ \bibinfo {pages} {1095} (\bibinfo {year}
  {1967})}\BibitemShut {NoStop}%
\bibitem [{\citenamefont {Ghizzo}\ \emph {et~al.}(1987)\citenamefont {Ghizzo},
  \citenamefont {Izrar}, \citenamefont {Bertrand}, \citenamefont {Feix},
  \citenamefont {Fijalkow},\ and\ \citenamefont {Shoucri}}]{ghizzo1987bgk}%
  \BibitemOpen
  \bibfield  {author} {\bibinfo {author} {\bibnamefont {Ghizzo}, \bibfnamefont
  {A.}}, \bibinfo {author} {\bibnamefont {Izrar}, \bibfnamefont {B.}}, \bibinfo
  {author} {\bibnamefont {Bertrand}, \bibfnamefont {P.}}, \bibinfo {author}
  {\bibnamefont {Feix}, \bibfnamefont {M.}}, \bibinfo {author} {\bibnamefont
  {Fijalkow}, \bibfnamefont {E.}}, \ and\ \bibinfo {author} {\bibnamefont
  {Shoucri}, \bibfnamefont {M.}},\ }\href@noop {} {\bibfield  {journal}
  {\bibinfo  {journal} {Physics Letters A}\ }\textbf {\bibinfo {volume}
  {120}},\ \bibinfo {pages} {191} (\bibinfo {year} {1987})}\BibitemShut
  {NoStop}%
\bibitem [{\citenamefont {Hirota}(1971)}]{Hirota1971}%
  \BibitemOpen
  \bibfield  {author} {\bibinfo {author} {\bibnamefont {Hirota}, \bibfnamefont
  {R.}},\ }\href@noop {} {\bibfield  {journal} {\bibinfo  {journal} {Physical
  Review Letters}\ }\textbf {\bibinfo {volume} {27}},\ \bibinfo {pages} {1192}
  (\bibinfo {year} {1971})}\BibitemShut {NoStop}%
\bibitem [{\citenamefont {Hirota}(1972)}]{Hirota1972}%
  \BibitemOpen
  \bibfield  {author} {\bibinfo {author} {\bibnamefont {Hirota}, \bibfnamefont
  {R.}},\ }\href@noop {} {\bibfield  {journal} {\bibinfo  {journal} {Journal of
  the Physical Society of Japan}\ }\textbf {\bibinfo {volume} {33}},\ \bibinfo
  {pages} {1456} (\bibinfo {year} {1972})}\BibitemShut {NoStop}%
\bibitem [{\citenamefont {Hobara}\ \emph {et~al.}(2008)\citenamefont {Hobara},
  \citenamefont {Walker}, \citenamefont {Balikhin}, \citenamefont {Pokhotelov},
  \citenamefont {Gedalin}, \citenamefont {Krasnoselskikh}, \citenamefont
  {Hayakawa}, \citenamefont {Andr{\'e}}, \citenamefont {Dunlop}, \citenamefont
  {R{\`e}me} \emph {et~al.}}]{hobara2008cluster}%
  \BibitemOpen
  \bibfield  {author} {\bibinfo {author} {\bibnamefont {Hobara}, \bibfnamefont
  {Y.}}, \bibinfo {author} {\bibnamefont {Walker}, \bibfnamefont {S.}},
  \bibinfo {author} {\bibnamefont {Balikhin}, \bibfnamefont {M.}}, \bibinfo
  {author} {\bibnamefont {Pokhotelov}, \bibfnamefont {O.}}, \bibinfo {author}
  {\bibnamefont {Gedalin}, \bibfnamefont {M.}}, \bibinfo {author} {\bibnamefont
  {Krasnoselskikh}, \bibfnamefont {V.}}, \bibinfo {author} {\bibnamefont
  {Hayakawa}, \bibfnamefont {M.}}, \bibinfo {author} {\bibnamefont {Andr{\'e}},
  \bibfnamefont {M.}}, \bibinfo {author} {\bibnamefont {Dunlop}, \bibfnamefont
  {M.}}, \bibinfo {author} {\bibnamefont {R{\`e}me}, \bibfnamefont {H.}},
  \emph {et~al.},\ }\href@noop {} {\bibfield  {journal} {\bibinfo  {journal}
  {Journal of Geophysical Research: Space Physics}\ }\textbf {\bibinfo {volume}
  {113}} (\bibinfo {year} {2008})}\BibitemShut {NoStop}%
\bibitem [{\citenamefont {Ikezi}, \citenamefont {Taylor},\ and\ \citenamefont
  {Baker}(1970)}]{ikezi1970formation}%
  \BibitemOpen
  \bibfield  {author} {\bibinfo {author} {\bibnamefont {Ikezi}, \bibfnamefont
  {H.}}, \bibinfo {author} {\bibnamefont {Taylor}, \bibfnamefont {R.}}, \ and\
  \bibinfo {author} {\bibnamefont {Baker}, \bibfnamefont {D.}},\ }\href@noop {}
  {\bibfield  {journal} {\bibinfo  {journal} {Physical Review Letters}\
  }\textbf {\bibinfo {volume} {25}},\ \bibinfo {pages} {11} (\bibinfo {year}
  {1970})}\BibitemShut {NoStop}%
\bibitem [{\citenamefont {Jenab}\ and\ \citenamefont
  {Kourakis}(2014{\natexlab{a}})}]{jenab2014multicomponent}%
  \BibitemOpen
  \bibfield  {author} {\bibinfo {author} {\bibnamefont {Jenab}, \bibfnamefont
  {S.~H.}}\ and\ \bibinfo {author} {\bibnamefont {Kourakis}, \bibfnamefont
  {I.}},\ }\href@noop {} {\bibfield  {journal} {\bibinfo  {journal} {Physics of
  Plasmas}\ }\textbf {\bibinfo {volume} {21}},\ \bibinfo {pages} {043701}
  (\bibinfo {year} {2014}{\natexlab{a}})}\BibitemShut {NoStop}%
\bibitem [{\citenamefont {Jenab}\ and\ \citenamefont
  {Kourakis}(2014{\natexlab{b}})}]{jenab2014vlasov}%
  \BibitemOpen
  \bibfield  {author} {\bibinfo {author} {\bibnamefont {Jenab}, \bibfnamefont
  {S.~M.~H.}}\ and\ \bibinfo {author} {\bibnamefont {Kourakis}, \bibfnamefont
  {I.}},\ }\href@noop {} {\bibfield  {journal} {\bibinfo  {journal} {The
  European Physical Journal D}\ }\textbf {\bibinfo {volume} {68}},\ \bibinfo
  {pages} {1} (\bibinfo {year} {2014}{\natexlab{b}})}\BibitemShut {NoStop}%
\bibitem [{\citenamefont {Kakad}\ \emph {et~al.}(2016)\citenamefont {Kakad},
  \citenamefont {Kakad}, \citenamefont {Anekallu}, \citenamefont {Lakhina},
  \citenamefont {Omura},\ and\ \citenamefont {Fazakerley}}]{kakad2016slow}%
  \BibitemOpen
  \bibfield  {author} {\bibinfo {author} {\bibnamefont {Kakad}, \bibfnamefont
  {A.}}, \bibinfo {author} {\bibnamefont {Kakad}, \bibfnamefont {B.}}, \bibinfo
  {author} {\bibnamefont {Anekallu}, \bibfnamefont {C.}}, \bibinfo {author}
  {\bibnamefont {Lakhina}, \bibfnamefont {G.}}, \bibinfo {author} {\bibnamefont
  {Omura}, \bibfnamefont {Y.}}, \ and\ \bibinfo {author} {\bibnamefont
  {Fazakerley}, \bibfnamefont {A.}},\ }\href@noop {} {\bibfield  {journal}
  {\bibinfo  {journal} {Journal of Geophysical Research: Space Physics}\
  }\textbf {\bibinfo {volume} {121}},\ \bibinfo {pages} {4452} (\bibinfo {year}
  {2016})}\BibitemShut {NoStop}%
\bibitem [{\citenamefont {Kakad}, \citenamefont {Omura},\ and\ \citenamefont
  {Kakad}(2013)}]{Kakad2013}%
  \BibitemOpen
  \bibfield  {author} {\bibinfo {author} {\bibnamefont {Kakad}, \bibfnamefont
  {A.}}, \bibinfo {author} {\bibnamefont {Omura}, \bibfnamefont {Y.}}, \ and\
  \bibinfo {author} {\bibnamefont {Kakad}, \bibfnamefont {B.}},\ }\href@noop {}
  {\bibfield  {journal} {\bibinfo  {journal} {Physics of Plasmas}\ }\textbf
  {\bibinfo {volume} {20}},\ \bibinfo {pages} {062103} (\bibinfo {year}
  {2013})}\BibitemShut {NoStop}%
\bibitem [{\citenamefont {Kakad}, \citenamefont {Kakad},\ and\ \citenamefont
  {Omura}(2014)}]{Kakad20145589}%
  \BibitemOpen
  \bibfield  {author} {\bibinfo {author} {\bibnamefont {Kakad}, \bibfnamefont
  {B.}}, \bibinfo {author} {\bibnamefont {Kakad}, \bibfnamefont {A.}}, \ and\
  \bibinfo {author} {\bibnamefont {Omura}, \bibfnamefont {Y.}},\ }\href@noop {}
  {\bibfield  {journal} {\bibinfo  {journal} {Journal of Geophysical Research:
  Space Physics}\ }\textbf {\bibinfo {volume} {119}},\ \bibinfo {pages} {5589}
  (\bibinfo {year} {2014})}\BibitemShut {NoStop}%
\bibitem [{\citenamefont {Kazeminezhad}, \citenamefont {Kuhn},\ and\
  \citenamefont {Tavakoli}(2003)}]{kazeminezhad2003vlasov}%
  \BibitemOpen
  \bibfield  {author} {\bibinfo {author} {\bibnamefont {Kazeminezhad},
  \bibfnamefont {F.}}, \bibinfo {author} {\bibnamefont {Kuhn}, \bibfnamefont
  {S.}}, \ and\ \bibinfo {author} {\bibnamefont {Tavakoli}, \bibfnamefont
  {A.}},\ }\href@noop {} {\bibfield  {journal} {\bibinfo  {journal} {Physical
  Review E}\ }\textbf {\bibinfo {volume} {67}},\ \bibinfo {pages} {026704}
  (\bibinfo {year} {2003})}\BibitemShut {NoStop}%
\bibitem [{\citenamefont {Kojima}\ \emph {et~al.}(1997)\citenamefont {Kojima},
  \citenamefont {Matsumoto}, \citenamefont {Chikuba}, \citenamefont {Horiyama},
  \citenamefont {Ashour-Abdalla},\ and\ \citenamefont
  {Anderson}}]{kojima1997geotail}%
  \BibitemOpen
  \bibfield  {author} {\bibinfo {author} {\bibnamefont {Kojima}, \bibfnamefont
  {H.}}, \bibinfo {author} {\bibnamefont {Matsumoto}, \bibfnamefont {H.}},
  \bibinfo {author} {\bibnamefont {Chikuba}, \bibfnamefont {S.}}, \bibinfo
  {author} {\bibnamefont {Horiyama}, \bibfnamefont {S.}}, \bibinfo {author}
  {\bibnamefont {Ashour-Abdalla}, \bibfnamefont {M.}}, \ and\ \bibinfo {author}
  {\bibnamefont {Anderson}, \bibfnamefont {R.}},\ }\href@noop {} {\bibfield
  {journal} {\bibinfo  {journal} {Journal of Geophysical Research: Space
  Physics}\ }\textbf {\bibinfo {volume} {102}},\ \bibinfo {pages} {14439}
  (\bibinfo {year} {1997})}\BibitemShut {NoStop}%
\bibitem [{\citenamefont {Lakhina}\ \emph {et~al.}(2009)\citenamefont
  {Lakhina}, \citenamefont {Singh}, \citenamefont {Kakad}, \citenamefont
  {Goldstein}, \citenamefont {Vinas},\ and\ \citenamefont
  {Pickett}}]{lakhina2009mechanism}%
  \BibitemOpen
  \bibfield  {author} {\bibinfo {author} {\bibnamefont {Lakhina}, \bibfnamefont
  {G.}}, \bibinfo {author} {\bibnamefont {Singh}, \bibfnamefont {S.}}, \bibinfo
  {author} {\bibnamefont {Kakad}, \bibfnamefont {A.}}, \bibinfo {author}
  {\bibnamefont {Goldstein}, \bibfnamefont {M.}}, \bibinfo {author}
  {\bibnamefont {Vinas}, \bibfnamefont {A.}}, \ and\ \bibinfo {author}
  {\bibnamefont {Pickett}, \bibfnamefont {J.}},\ }\href@noop {} {\bibfield
  {journal} {\bibinfo  {journal} {Journal of Geophysical Research: Space
  Physics}\ }\textbf {\bibinfo {volume} {114}} (\bibinfo {year}
  {2009})}\BibitemShut {NoStop}%
\bibitem [{\citenamefont {Lee}\ and\ \citenamefont
  {Sakthivel}(2011)}]{lee2011exact}%
  \BibitemOpen
  \bibfield  {author} {\bibinfo {author} {\bibnamefont {Lee}, \bibfnamefont
  {J.}}\ and\ \bibinfo {author} {\bibnamefont {Sakthivel}, \bibfnamefont
  {R.}},\ }\href@noop {} {\bibfield  {journal} {\bibinfo  {journal} {Reports on
  Mathematical Physics}\ }\textbf {\bibinfo {volume} {68}},\ \bibinfo {pages}
  {153} (\bibinfo {year} {2011})}\BibitemShut {NoStop}%
\bibitem [{\citenamefont {Lonngren}(1998)}]{lonngren1998ion}%
  \BibitemOpen
  \bibfield  {author} {\bibinfo {author} {\bibnamefont {Lonngren},
  \bibfnamefont {K.~E.}},\ }\href@noop {} {\bibfield  {journal} {\bibinfo
  {journal} {Optical and quantum electronics}\ }\textbf {\bibinfo {volume}
  {30}},\ \bibinfo {pages} {615} (\bibinfo {year} {1998})}\BibitemShut
  {NoStop}%
\bibitem [{\citenamefont {Ludwig}, \citenamefont {Ferreira},\ and\
  \citenamefont {Nakamura}(1984)}]{ludwig1984observation}%
  \BibitemOpen
  \bibfield  {author} {\bibinfo {author} {\bibnamefont {Ludwig}, \bibfnamefont
  {G.}}, \bibinfo {author} {\bibnamefont {Ferreira}, \bibfnamefont {J.}}, \
  and\ \bibinfo {author} {\bibnamefont {Nakamura}, \bibfnamefont {Y.}},\
  }\href@noop {} {\bibfield  {journal} {\bibinfo  {journal} {Physical review
  letters}\ }\textbf {\bibinfo {volume} {52}},\ \bibinfo {pages} {275}
  (\bibinfo {year} {1984})}\BibitemShut {NoStop}%
\bibitem [{\citenamefont {Matsumoto}\ \emph {et~al.}(1994)\citenamefont
  {Matsumoto}, \citenamefont {Kojima}, \citenamefont {Miyatake}, \citenamefont
  {Omura}, \citenamefont {Okada}, \citenamefont {Nagano},\ and\ \citenamefont
  {Tsutsui}}]{matsumoto1994electrostatic}%
  \BibitemOpen
  \bibfield  {author} {\bibinfo {author} {\bibnamefont {Matsumoto},
  \bibfnamefont {H.}}, \bibinfo {author} {\bibnamefont {Kojima}, \bibfnamefont
  {H.}}, \bibinfo {author} {\bibnamefont {Miyatake}, \bibfnamefont {T.}},
  \bibinfo {author} {\bibnamefont {Omura}, \bibfnamefont {Y.}}, \bibinfo
  {author} {\bibnamefont {Okada}, \bibfnamefont {M.}}, \bibinfo {author}
  {\bibnamefont {Nagano}, \bibfnamefont {I.}}, \ and\ \bibinfo {author}
  {\bibnamefont {Tsutsui}, \bibfnamefont {M.}},\ }\href@noop {} {\bibfield
  {journal} {\bibinfo  {journal} {Geophysical Research Letters}\ }\textbf
  {\bibinfo {volume} {21}},\ \bibinfo {pages} {2915} (\bibinfo {year}
  {1994})}\BibitemShut {NoStop}%
\bibitem [{\citenamefont {Nakamura}, \citenamefont {Ferreira},\ and\
  \citenamefont {Ludwig}(1985)}]{nakamura1985experiments}%
  \BibitemOpen
  \bibfield  {author} {\bibinfo {author} {\bibnamefont {Nakamura},
  \bibfnamefont {Y.}}, \bibinfo {author} {\bibnamefont {Ferreira},
  \bibfnamefont {J.}}, \ and\ \bibinfo {author} {\bibnamefont {Ludwig},
  \bibfnamefont {G.}},\ }\href@noop {} {\bibfield  {journal} {\bibinfo
  {journal} {Journal of plasma physics}\ }\textbf {\bibinfo {volume} {33}},\
  \bibinfo {pages} {237} (\bibinfo {year} {1985})}\BibitemShut {NoStop}%
\bibitem [{\citenamefont {Nunn}(1993)}]{nunn1993novel}%
  \BibitemOpen
  \bibfield  {author} {\bibinfo {author} {\bibnamefont {Nunn}, \bibfnamefont
  {D.}},\ }\href@noop {} {\bibfield  {journal} {\bibinfo  {journal} {Journal of
  Computational Physics}\ }\textbf {\bibinfo {volume} {108}},\ \bibinfo {pages}
  {180} (\bibinfo {year} {1993})}\BibitemShut {NoStop}%
\bibitem [{\citenamefont {Pickett}\ \emph {et~al.}(2004)\citenamefont
  {Pickett}, \citenamefont {Kahler}, \citenamefont {Chen}, \citenamefont
  {Huff}, \citenamefont {Santolik}, \citenamefont {Khotyaintsev}, \citenamefont
  {D{\'e}cr{\'e}au}, \citenamefont {Winningham}, \citenamefont {Frahm},
  \citenamefont {Goldstein} \emph {et~al.}}]{pickett2004solitary}%
  \BibitemOpen
  \bibfield  {author} {\bibinfo {author} {\bibnamefont {Pickett}, \bibfnamefont
  {J.}}, \bibinfo {author} {\bibnamefont {Kahler}, \bibfnamefont {S.}},
  \bibinfo {author} {\bibnamefont {Chen}, \bibfnamefont {L.-J.}}, \bibinfo
  {author} {\bibnamefont {Huff}, \bibfnamefont {R.}}, \bibinfo {author}
  {\bibnamefont {Santolik}, \bibfnamefont {O.}}, \bibinfo {author}
  {\bibnamefont {Khotyaintsev}, \bibfnamefont {Y.}}, \bibinfo {author}
  {\bibnamefont {D{\'e}cr{\'e}au}, \bibfnamefont {P.}}, \bibinfo {author}
  {\bibnamefont {Winningham}, \bibfnamefont {D.}}, \bibinfo {author}
  {\bibnamefont {Frahm}, \bibfnamefont {R.}}, \bibinfo {author} {\bibnamefont
  {Goldstein}, \bibfnamefont {M.}},  \emph {et~al.},\ }\href@noop {} {\bibfield
   {journal} {\bibinfo  {journal} {Nonlinear Processes in Geophysics}\ }\textbf
  {\bibinfo {volume} {11}},\ \bibinfo {pages} {183} (\bibinfo {year}
  {2004})}\BibitemShut {NoStop}%
\bibitem [{\citenamefont {Pickett}\ \emph {et~al.}(2003)\citenamefont
  {Pickett}, \citenamefont {Menietti}, \citenamefont {Gurnett}, \citenamefont
  {Tsurutani}, \citenamefont {Kintner}, \citenamefont {Klatt},\ and\
  \citenamefont {Balogh}}]{pickett2003solitary}%
  \BibitemOpen
  \bibfield  {author} {\bibinfo {author} {\bibnamefont {Pickett}, \bibfnamefont
  {J.}}, \bibinfo {author} {\bibnamefont {Menietti}, \bibfnamefont {J.}},
  \bibinfo {author} {\bibnamefont {Gurnett}, \bibfnamefont {D.}}, \bibinfo
  {author} {\bibnamefont {Tsurutani}, \bibfnamefont {B.}}, \bibinfo {author}
  {\bibnamefont {Kintner}, \bibfnamefont {P.}}, \bibinfo {author} {\bibnamefont
  {Klatt}, \bibfnamefont {E.}}, \ and\ \bibinfo {author} {\bibnamefont
  {Balogh}, \bibfnamefont {A.}},\ }\href@noop {} {\bibfield  {journal}
  {\bibinfo  {journal} {Nonlinear Processes in Geophysics}\ }\textbf {\bibinfo
  {volume} {10}},\ \bibinfo {pages} {3} (\bibinfo {year} {2003})}\BibitemShut
  {NoStop}%
\bibitem [{\citenamefont {Qi}\ \emph {et~al.}(2015)\citenamefont {Qi},
  \citenamefont {Xu}, \citenamefont {Zhao}, \citenamefont {Zhang},
  \citenamefont {Duan},\ and\ \citenamefont {Yang}}]{Qi20153815}%
  \BibitemOpen
  \bibfield  {author} {\bibinfo {author} {\bibnamefont {Qi}, \bibfnamefont
  {X.}}, \bibinfo {author} {\bibnamefont {Xu}, \bibfnamefont {Y.-X.}}, \bibinfo
  {author} {\bibnamefont {Zhao}, \bibfnamefont {X.-Y.}}, \bibinfo {author}
  {\bibnamefont {Zhang}, \bibfnamefont {L.-Y.}}, \bibinfo {author}
  {\bibnamefont {Duan}, \bibfnamefont {W.-S.}}, \ and\ \bibinfo {author}
  {\bibnamefont {Yang}, \bibfnamefont {L.}},\ }\href@noop {} {\bibfield
  {journal} {\bibinfo  {journal} {IEEE Transactions on Plasma Science}\
  }\textbf {\bibinfo {volume} {43}},\ \bibinfo {pages} {3815} (\bibinfo {year}
  {2015})}\BibitemShut {NoStop}%
\bibitem [{\citenamefont {Schamel}(1971)}]{schamel_1}%
  \BibitemOpen
  \bibfield  {author} {\bibinfo {author} {\bibnamefont {Schamel}, \bibfnamefont
  {H.}},\ }\href@noop {} {\bibfield  {journal} {\bibinfo  {journal} {Plasma
  Physics}\ }\textbf {\bibinfo {volume} {13}},\ \bibinfo {pages} {491}
  (\bibinfo {year} {1971})}\BibitemShut {NoStop}%
\bibitem [{\citenamefont {Schamel}(1972{\natexlab{a}})}]{schamel_2}%
  \BibitemOpen
  \bibfield  {author} {\bibinfo {author} {\bibnamefont {Schamel}, \bibfnamefont
  {H.}},\ }\href@noop {} {\bibfield  {journal} {\bibinfo  {journal} {Journal of
  Plasma Physics}\ }\textbf {\bibinfo {volume} {7}},\ \bibinfo {pages} {1}
  (\bibinfo {year} {1972}{\natexlab{a}})}\BibitemShut {NoStop}%
\bibitem [{\citenamefont {Schamel}(1972{\natexlab{b}})}]{schamel_3}%
  \BibitemOpen
  \bibfield  {author} {\bibinfo {author} {\bibnamefont {Schamel}, \bibfnamefont
  {H.}},\ }\href@noop {} {\bibfield  {journal} {\bibinfo  {journal} {Plasma
  Physics}\ }\textbf {\bibinfo {volume} {14}},\ \bibinfo {pages} {905}
  (\bibinfo {year} {1972}{\natexlab{b}})}\BibitemShut {NoStop}%
\bibitem [{\citenamefont {Schamel}(1973)}]{schamel_4}%
  \BibitemOpen
  \bibfield  {author} {\bibinfo {author} {\bibnamefont {Schamel}, \bibfnamefont
  {H.}},\ }\href@noop {} {\bibfield  {journal} {\bibinfo  {journal} {Journal of
  Plasma Physics}\ }\textbf {\bibinfo {volume} {9}},\ \bibinfo {pages} {377}
  (\bibinfo {year} {1973})}\BibitemShut {NoStop}%
\bibitem [{\citenamefont {Schamel}(1979)}]{schamel_5}%
  \BibitemOpen
  \bibfield  {author} {\bibinfo {author} {\bibnamefont {Schamel}, \bibfnamefont
  {H.}},\ }\href@noop {} {\bibfield  {journal} {\bibinfo  {journal} {Physica
  Scripta}\ }\textbf {\bibinfo {volume} {20}},\ \bibinfo {pages} {306}
  (\bibinfo {year} {1979})}\BibitemShut {NoStop}%
\bibitem [{\citenamefont {Sharma}, \citenamefont {Sengupta},\ and\
  \citenamefont {Sen}(2015)}]{Sharma2015}%
  \BibitemOpen
  \bibfield  {author} {\bibinfo {author} {\bibnamefont {Sharma}, \bibfnamefont
  {S.}}, \bibinfo {author} {\bibnamefont {Sengupta}, \bibfnamefont {S.}}, \
  and\ \bibinfo {author} {\bibnamefont {Sen}, \bibfnamefont {A.}},\ }\href@noop
  {} {\bibfield  {journal} {\bibinfo  {journal} {Physics of Plasmas}\ }\textbf
  {\bibinfo {volume} {22}},\ \bibinfo {pages} {022115} (\bibinfo {year}
  {2015})}\BibitemShut {NoStop}%
\bibitem [{\citenamefont {Sultana}\ and\ \citenamefont
  {Kourakis}(2015)}]{sultana2015electron}%
  \BibitemOpen
  \bibfield  {author} {\bibinfo {author} {\bibnamefont {Sultana}, \bibfnamefont
  {S.}}\ and\ \bibinfo {author} {\bibnamefont {Kourakis}, \bibfnamefont {I.}},\
  }\href@noop {} {\bibfield  {journal} {\bibinfo  {journal} {Physics of
  Plasmas}\ }\textbf {\bibinfo {volume} {22}},\ \bibinfo {pages} {102302}
  (\bibinfo {year} {2015})}\BibitemShut {NoStop}%
\bibitem [{\citenamefont {Taha}, \citenamefont {Noorani},\ and\ \citenamefont
  {Hashim}(2013)}]{taha2013new}%
  \BibitemOpen
  \bibfield  {author} {\bibinfo {author} {\bibnamefont {Taha}, \bibfnamefont
  {W.~M.}}, \bibinfo {author} {\bibnamefont {Noorani}, \bibfnamefont
  {M.~S.~M.}}, \ and\ \bibinfo {author} {\bibnamefont {Hashim}, \bibfnamefont
  {I.}},\ }\href@noop {} {\bibfield  {journal} {\bibinfo  {journal} {Journal of
  Applied Mathematics}\ }\textbf {\bibinfo {volume} {2013}} (\bibinfo {year}
  {2013})}\BibitemShut {NoStop}%
\bibitem [{\citenamefont {Verheest}(1988)}]{verheest1988ion}%
  \BibitemOpen
  \bibfield  {author} {\bibinfo {author} {\bibnamefont {Verheest},
  \bibfnamefont {F.}},\ }\href@noop {} {\bibfield  {journal} {\bibinfo
  {journal} {Journal of plasma physics}\ }\textbf {\bibinfo {volume} {39}},\
  \bibinfo {pages} {71} (\bibinfo {year} {1988})}\BibitemShut {NoStop}%
\bibitem [{\citenamefont {Verheest}, \citenamefont {Hellberg},\ and\
  \citenamefont {Kourakis}(2013)}]{Verheest2013}%
  \BibitemOpen
  \bibfield  {author} {\bibinfo {author} {\bibnamefont {Verheest},
  \bibfnamefont {F.}}, \bibinfo {author} {\bibnamefont {Hellberg},
  \bibfnamefont {M.~A.}}, \ and\ \bibinfo {author} {\bibnamefont {Kourakis},
  \bibfnamefont {I.}},\ }\href@noop {} {\bibfield  {journal} {\bibinfo
  {journal} {Physics of Plasmas}\ }\textbf {\bibinfo {volume} {20}},\ \bibinfo
  {pages} {012302} (\bibinfo {year} {2013})}\BibitemShut {NoStop}%
\bibitem [{\citenamefont {Verheest}, \citenamefont {Olivier},\ and\
  \citenamefont {Hereman}(2016)}]{verheest2016modified}%
  \BibitemOpen
  \bibfield  {author} {\bibinfo {author} {\bibnamefont {Verheest},
  \bibfnamefont {F.}}, \bibinfo {author} {\bibnamefont {Olivier}, \bibfnamefont
  {C.~P.}}, \ and\ \bibinfo {author} {\bibnamefont {Hereman}, \bibfnamefont
  {W.~A.}},\ }\href@noop {} {\bibfield  {journal} {\bibinfo  {journal} {arXiv
  preprint arXiv:1604.03097}\ } (\bibinfo {year} {2016})}\BibitemShut {NoStop}%
\bibitem [{\citenamefont {Wadati}(2001)}]{Wadati2001841}%
  \BibitemOpen
  \bibfield  {author} {\bibinfo {author} {\bibnamefont {Wadati}, \bibfnamefont
  {M.}},\ }\href@noop {} {\bibfield  {journal} {\bibinfo  {journal} {Pramana}\
  }\textbf {\bibinfo {volume} {57}},\ \bibinfo {pages} {841} (\bibinfo {year}
  {2001})}\BibitemShut {NoStop}%
\bibitem [{\citenamefont {Washimi}\ and\ \citenamefont
  {Taniuti}(1966)}]{Washimi1966996}%
  \BibitemOpen
  \bibfield  {author} {\bibinfo {author} {\bibnamefont {Washimi}, \bibfnamefont
  {H.}}\ and\ \bibinfo {author} {\bibnamefont {Taniuti}, \bibfnamefont {T.}},\
  }\href@noop {} {\bibfield  {journal} {\bibinfo  {journal} {Physical Review
  Letters}\ }\textbf {\bibinfo {volume} {17}},\ \bibinfo {pages} {996}
  (\bibinfo {year} {1966})}\BibitemShut {NoStop}%
\bibitem [{\citenamefont {Wu}\ and\ \citenamefont {Liu}(2013)}]{wu2013new}%
  \BibitemOpen
  \bibfield  {author} {\bibinfo {author} {\bibnamefont {Wu}, \bibfnamefont
  {Y.}}\ and\ \bibinfo {author} {\bibnamefont {Liu}, \bibfnamefont {Z.}},\ }in\
  \href@noop {} {\emph {\bibinfo {booktitle} {Abstract and Applied
  Analysis}}},\ Vol.\ \bibinfo {volume} {2013}\ (\bibinfo {organization}
  {Hindawi Publishing Corporation},\ \bibinfo {year} {2013})\BibitemShut
  {NoStop}%
\bibitem [{\citenamefont {Zabusky}\ and\ \citenamefont
  {Kruskal}(1965)}]{Zabusky1965}%
  \BibitemOpen
  \bibfield  {author} {\bibinfo {author} {\bibnamefont {Zabusky}, \bibfnamefont
  {N.~J.}}\ and\ \bibinfo {author} {\bibnamefont {Kruskal}, \bibfnamefont
  {M.~D.}},\ }\href@noop {} {\bibfield  {journal} {\bibinfo  {journal}
  {Physical review letters}\ }\textbf {\bibinfo {volume} {15}},\ \bibinfo
  {pages} {240} (\bibinfo {year} {1965})}\BibitemShut {NoStop}%
\end{thebibliography}%

%

\end{document}